\documentclass[aps,prb,10pt,amsmath,amssymb,twocolumn,footinbib,superscriptaddress]{revtex4-1}

\usepackage{color, braket}
\usepackage[large,bf]{subfigure}     

\usepackage[breaklinks]{hyperref}
\newcommand{\Z}[1]{\texorpdfstring{\ensuremath{\mathbb{Z}_{#1}}}{Z#1}}

\definecolor{purple}{rgb}{0.5,0,0.5}
\definecolor{dkgreen}{rgb}{0,0.5,0}
\definecolor{orange}{rgb}{1,0.5,0}

\usepackage{ifpdf}
\ifpdf
    \usepackage{psfrag}
    \usepackage[pdftex]{epsfig}
\else
    \usepackage{psfrag}
    \usepackage[dvips]{epsfig}
\fi

\begin{document}

\title{Assembling Fibonacci Anyons From a \Z3 Parafermion Lattice Model}
\date{\today}
\author{E. M. Stoudenmire}
\affiliation{Perimeter Institute of Theoretical Physics, Waterloo, Ontario, N2L 2Y5, Canada}
\author{David J. Clarke}
\affiliation{Department of Physics and Institute for Quantum Information and Matter, California Institute of Technology, Pasadena, CA 91125, USA}
\affiliation{Microsoft Research, Station Q, University of California, Santa Barbara, CA 93106, USA}
\affiliation{Condensed Matter Theory Center, Department of Physics, University of Maryland, College Park, Maryland 20742, USA}
\author{Roger S. K. Mong}
\affiliation{Department of Physics and Institute for Quantum Information and Matter, California Institute of Technology, Pasadena, CA 91125, USA}
\affiliation{Walter Burke Institute for Theoretical Physics, California Institute of Technology, Pasadena, CA 91125, USA}
\affiliation{Department of Physics and Astronomy, University of Pittsburgh, Pittsburgh, PA 15260, USA}
\author{Jason Alicea}
\affiliation{Department of Physics and Institute for Quantum Information and Matter, California Institute of Technology, Pasadena, CA 91125, USA}
\affiliation{Walter Burke Institute for Theoretical Physics, California Institute of Technology, Pasadena, CA 91125, USA}


\begin{abstract}
Recent concrete proposals suggest it is possible to engineer a two-dimensional bulk phase supporting non-Abelian Fibonacci anyons
out of Abelian fractional quantum Hall systems.
The low-energy degrees of freedom of such setups can be modeled as $\Z3$ parafermions ``hopping'' on a two-dimensional lattice.
We use the density matrix renormalization group to study a model of this type interpolating between the decoupled-chain,
triangular-lattice, and square-lattice limits. The results show clear evidence of the Fibonacci phase over a wide region of the phase diagram,
most notably including the isotropic triangular lattice point. We also study the broader phase diagram of this model and show that elsewhere it supports an Abelian state with semionic excitations.
\end{abstract}

\maketitle

\section{Introduction}

The experimental realization of non-Abelian anyons---emergent particles with highly exotic exchange statistics---is anticipated to have widespread implications.
Apart from demonstrating a profound new facet of quantum mechanics, many applications await the completion of this ongoing quest.
Examples include tests of Bell's inequalities,\cite{Drummond:2014} robust quantum memory, novel low-temperature circuit elements,\cite{Clarke:2014b} and most importantly intrinsically fault-tolerant `topological' quantum computing.\cite{Kitaev:2003,Freedman:1998,Freedman:2003,Bonderson:2008,Nayak:2008}
The last of these relies on the remarkable fact that adiabatic exchange (braiding) of non-Abelian anyons enacts a unitary rotation within the space of locally indistinguishable ground states generated by the anyons.
Storage and manipulation of qubits encoded in these ground states thus takes place non-locally, so that the quantum information is securely `hidden' from local environmental perturbations.

While most of these applications can be carried out with any non-Abelian anyon type, topological quantum computation carries more stringent demands.
Consider Ising anyons---or defects that bind Majorana zero modes---which likely constitute the most experimentally accessible non-Abelian anyon.
Numerous realistic platforms now exist for realizing Ising anyons, most prominently in quantum Hall systems \cite{Moore:1991,Nayak:2008} and topological superconductors,\cite{Read:2000,Kitaev:2001,Beenakker:2013,Alicea:2012n,Leijnse:2012,Stanescu:2013,Elliot:2014} and indeed tantalizing experimental evidence of these particles has accumulated in both settings.\cite{Radu:2008,Dolev:2008,Willett:2009,Willett:2010,An:2011,Willett:2013,Mourik:2012,Das:2012,Rokhinson:2012,Deng:2012,Finck:2013,Churchill:2013,Nadj-Perge}
Braiding Ising anyons, however, only amounts to 90$^\circ$ qubit rotations on the Bloch sphere.
Performing the arbitrary qubit rotations necessary for universal computing with Ising anyons requires additional operations that are not topologically protected.

This shortcoming strongly motivates the pursuit of other types of non-Abelian anyons with `denser' braid statistics.
One appealing alternative class are defects binding \emph{parafermionic zero modes},\cite{Fendley:2012} which comprise natural Majorana generalizations.
Although such defects require a strongly interacting host system (unlike Ising anyons), many plausible realizations have been suggested such as lattice quantum anomalous Hall states,\cite{Barkeshli:2012} Abelian quantum Hall/superconductor heterostructures,\cite{Lindner:2012,Clarke:2013a,Cheng:2012,Vaezi:2013,Klinovaja:2014} multilayer quantum Hall systems,\cite{Barkeshli:2013a,Barkeshli:2014b} and coupled-wire arrays.\cite{Oreg:2014,Klinovaja:2014a,Klinovaja:2014b}
Parafermionic zero modes produce a larger ground-state degeneracy than Majorana modes and thus enable a denser set of qubit rotations through braiding.
While providing some advantages for quantum computation,\cite{Clarke:2013a} their braid statistics nevertheless remains non-universal.\footnote{See also Ref.~\onlinecite{Hastings:2013} which discusses computational power in a related (though physically distinct) context.}

\begin{figure}[t]
\includegraphics[width=180px]{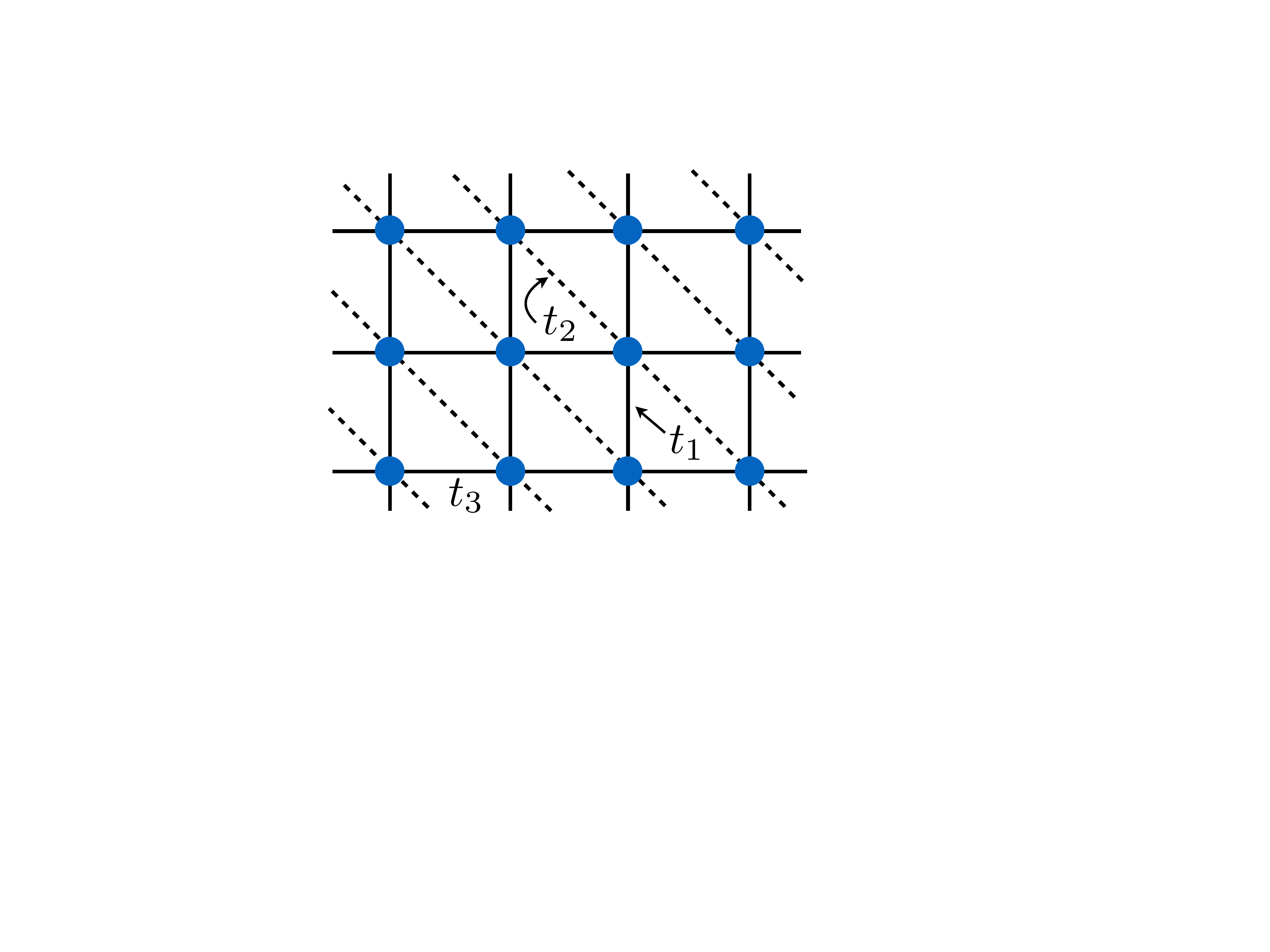}
\caption{Model of $\Z3$ parafermions with interactions forming an anisotropic triangular lattice with intrachain couplings $t_3$ (horizontal) and interchain couplings $t_1$ (vertical) and $t_2$ (slanted).}
\label{fig:triangular_lattice}
\end{figure}
Fortunately one can leverage setups supporting parafermionic zero modes to generate non-Abelian anyons allowing bona fide universal topological quantum computation.
To illustrate how this is possible, imagine nucleating a two-dimensional (2D) array of parafermionic zero modes in, say, a quantum Hall/superconductor hybrid structure.
The collection of zero modes encodes a \emph{macroscopic} ground-state degeneracy, similar to a partially filled Landau level.
Hybridizing parafermions among the sites can lift this degeneracy and produce new, possibly very exotic 2D phases.
Resolving the precise state selected by the coupled parafermion modes, however, poses a highly nontrivial task since the system can not be described by free particles (much like the Landau level problem with Coulomb interactions).

Inspired by the important work of Teo and Kane,\cite{Teo:2014} Ref.~\onlinecite{Mong:2014} nevertheless identified an analytically tractable limit
 where strongly anisotropic, weakly coupled parafermion chains could be shown to enter a topologically ordered `Fibonacci phase'.
This phase represents a cousin of the so-called $\Z3$ Read-Rezayi non-Abelian quantum Hall state,\cite{Read:1999} yet is built from well-understood Abelian states of matter.
(The analogy closely parallels the relation between a $p+ip$ superconductor and the Moore-Read state.\cite{Moore:1991,Read:2000})
Like its quantum Hall cousin, the Fibonacci phase supports Fibonacci anyons, which are the holy grail for topological quantum computation since they allow one to approximate arbitrary unitary gates to any desired accuracy solely through braiding.\cite{Nayak:2008,Freedman:2002a,Freedman:2002b}

While this progress is encouraging, the stability of the Fibonacci phase away from the solvable quasi-1D limit---as well as the broader phase diagram for coupled 2D parafermion arrays---remains largely unknown.
Obtaining a more thorough and quantitative understanding of these physical systems poses a pressing issue given the importance of finding experimentally accessible realizations of Fibonacci anyons for quantum computing.
In this paper we take a major step in this direction by performing extensive density matrix renormalization group (DMRG)\cite{Schollwoeck:2005,Schollwoeck:2011,ITensor} simulations of lattice parafermions with couplings sketched in Fig.~\ref{fig:triangular_lattice}.
A virtue of the model we study is that it interpolates between various physically interesting regimes, including decoupled chains and the isotropic triangular- and square-lattice limits as special cases.

DMRG naturally complements the weakly-coupled chain approach to realizing the Fibonacci phase. 
A key observation from analytics is that the `$y$' correlation length for the Fibonacci phase becomes arbitrarily small with weakly coupled chains, albeit at the expense of extended `$x$' correlations.
This tradeoff is highly favorable for numerics.
The short $y$ correlation length allows us to well-approximate the 2D limit of interest using systems composed of relatively few chains.
Moreover, with sufficient effort DMRG can handle very long chains in the $x$ direction even when each is close to criticality,
 facilitating direct comparisons to analytical predictions.
Near the decoupled-chain limit we can unambiguously pinpoint the Fibonacci phase in our simulations, then systematically track how it evolves as we gradually enhance the interchain coupling---all the while keeping the $y$ correlation length manageable.
The latter feature is important since in practice accurate DMRG results for typical 2D models on cylinders can only be obtained
for circumferences on the order of ten sites along $y$.
Finally, we note that moving away from the decoupled-chain limit lessens the overall entanglement, mitigating the cost of
the additional sites needed along the $y$ direction and allowing DMRG to even address the behavior of isotropic 2D systems.

In this way we  numerically show that the Fibonacci phase extends across a very wide swath of parameter space in our model, persisting from weakly coupled chains all the way to the isotropic triangular lattice point and beyond.
Evidently the quasi-1D limit pursued earlier is by no means necessary, but merely provides a convenient entryway into the relevant physics.
The broad stability window for Fibonacci anyons that we identify is one of the main punchlines of this paper.
We also present evidence for a second topologically ordered state that is Abelian and supports a semion as its only nontrivial quasiparticle.
This phase naturally emerges from weakly coupled chains upon swapping the sign for the intrachain coupling constant, and in our simulations also appears quite robust even away from this limit.
In fact the Fibonacci and semion phases comprise the only two states that appeared throughout the broad (though not completely exhaustive) parameter space that we explored numerically.
For a summary of our results see the octahedral phase diagram presented as a cutout template in Fig.~\ref{fig:phase_octahedron}.

We organize the remainder of the paper as follows.  Section~\ref{sec:models} motivates the lattice model we study from the viewpoint of physical quantum Hall-based architectures.
Section~\ref{sec:two_chain} analyzes the model, both analytically and numerically, in the two-chain limit---which we argue already contains precursors of Fibonacci physics.
The multi-chain setup of greatest interest is tackled in Sec.~\ref{sec:cylinders}.
We present a wealth of numerical evidence indicating the robustness of the Fibonacci phase; analytically capture the semion state; and explore the broader phase diagram of the model.
Section~\ref{Conclusions} discusses future directions and highlights the connection between our study and recent related works.
Four appendices contain additional details of our model and simulation methods.

\section{\Z3 Parafermion Lattice Model}\label{sec:models}

\begin{figure}
\includegraphics[width=230px]{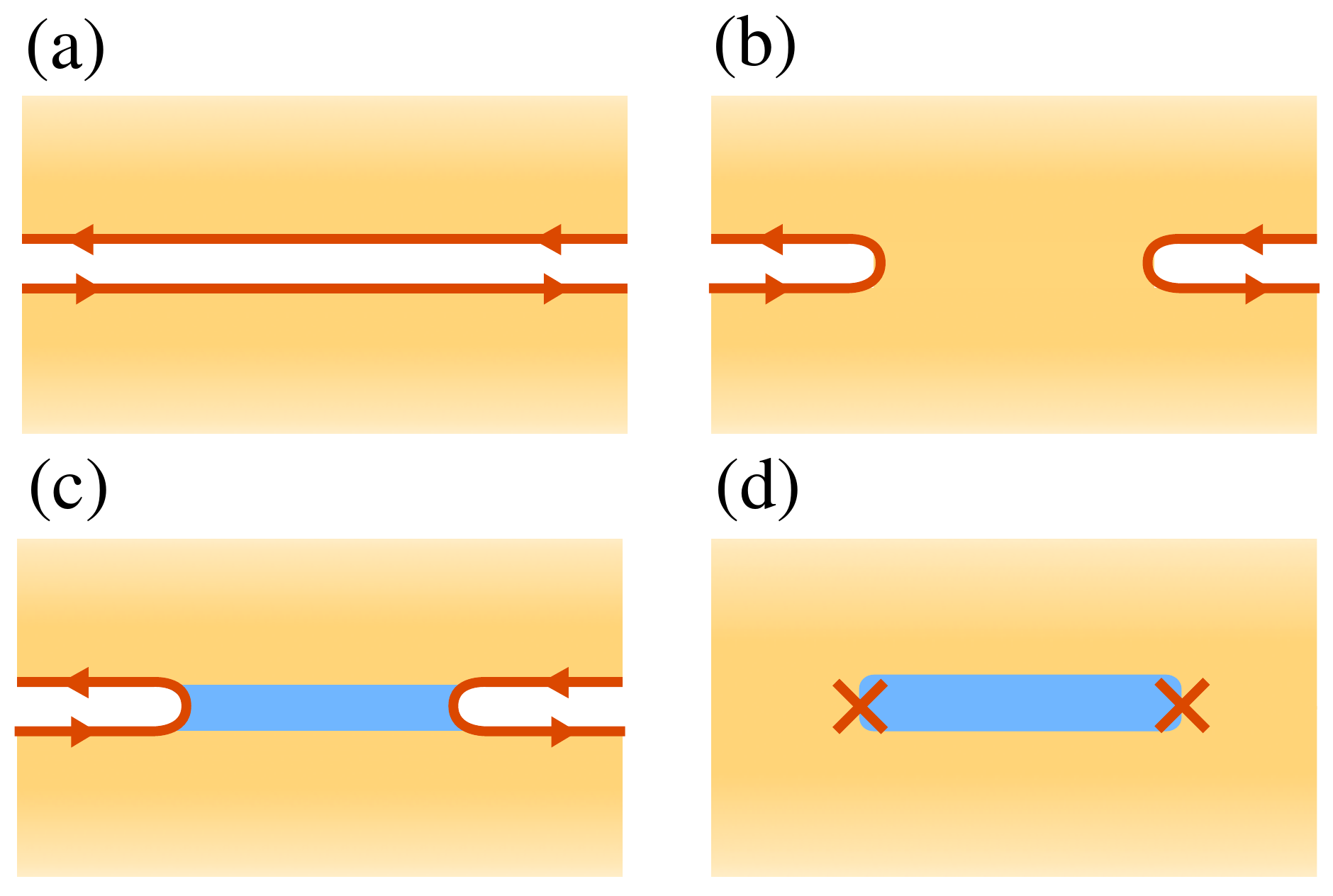}
\caption{(a) A trench dividing two regions of an Abelian quantum Hall phase supports gapless modes.
(b) Gapping these modes via electron backscattering recovers the original bulk phase.
(c) Introducing either Cooper pairing, or `crossed' tunneling in a bilayer system, provides an inequivalent way of gapping the trench.
(d) Domain walls between incompatibly gapped regions support protected zero modes.}
\label{fig:gapping}
\end{figure}

To motivate the lattice model we will study, let us review how a non-Abelian phase supporting bulk Fibonacci anyons can arise from coupling parafermionic zero modes nucleated in a fractional quantum Hall fluid.
As a primer consider some arbitrary Abelian quantum Hall state sliced into two adjacent halves as shown in Fig.~\ref{fig:gapping}(a).
The cut produces a new set of gapless counterpropagating edge states opposite the trench.
One can always fully gap these modes---effectively resewing the two halves---in more than one physically distinct way.
The most natural mechanism involves backscattering electrons across the trench to simply recover the original uninterrupted quantum Hall state as Fig.~\ref{fig:gapping}(b) illustrates.
Filling the trench with a superconductor [blue region in Fig.~\ref{fig:gapping}(c)] provides a second, intuitively quite different method: the edge modes can then gap out by Cooper pairing electrons from opposite sides of the trench.\cite{Lindner:2012,Clarke:2013a,Cheng:2012}
Some setups can support alternative charge-conserving gapping mechanisms (aside from trivial backscattering).\cite{Barkeshli:2013a,Barkeshli:2014b}
For instance, in quantum Hall bilayers formed out of Laughlin states at filling $\nu = 1/m$, one can gap the trench by `crossed' tunneling whereby electrons hop between the top layer on one side and the bottom layer on the other, sewing the halves in yet a different manner.
Domain walls [Fig.~\ref{fig:gapping}(d)] separating regions gapped in such incompatible ways bind protected zero-energy modes that form the basic building blocks in our lattice model.

\begin{figure}
\includegraphics[width=240px]{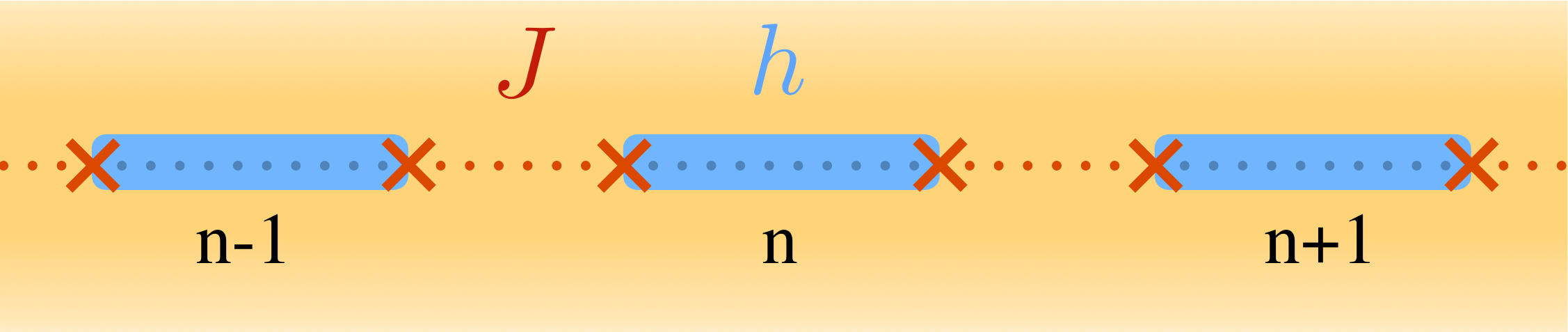}
\caption{One-dimensional chain of parafermionic zero modes arising from inequivalently gapped regions of a trench.  Hybridization among nearest-neighbors in the  chain is described by the Hamiltonian in Eq.~\eqref{eqn:single_trench}. }
\label{fig:single_trench}
\end{figure}

We are specifically interested in setups that support $\Z3$ parafermionic zero modes, which provide the minimal extension of the more familiar Majorana operators.
Concrete realizations include $(i)$ domain walls between pairing- and tunneling-gapped trenches in spin-unpolarized $\nu = 2/3$ quantum Hall fluids\cite{Mong:2014}  and $(ii)$ domain walls between trivial and `crossed' electron tunneling in $\nu = 1/3$ bilayers.\cite{Barkeshli:2014b}
Consider for the moment a one-dimensional array of well-separated $\Z3$ zero modes as in the geometry displayed in Fig.~\ref{fig:single_trench}.
These modes encode a ground-state degeneracy that can be understood as follows.
In the $\nu = 2/3$ realization each superconducting-gapped region can accommodate charge $2e/3$ without energy penalty, while in the bilayer example each `crossed' region can similarly acquire an $e/3$ dipole for free.

Parafermionic zero-mode operators cycle the system through the corresponding degenerate ground-state manifold.
Importantly, charges can be added to the trench in two physically distinct ways---and hence there exists two inequivalent representations of the parafermion operators.
We denote these two parafermion representations by $\alpha_{R,j}$ and $\alpha_{L,j}$ with $j$ the domain-wall site index.
As a specific example, in the $\nu = 2/3$ setup $\alpha_{L,j}$ and $\alpha_{R,j}$ alter the adjacent superconducting region by adding charge $2e/3$ to the upper and lower trench edges, respectively.
This distinction is meaningful since fractional charge can not migrate across the trench.
For further discussion on this important point see Ref.~\onlinecite{Mong:2014} as well as Appendix~\ref{appendix:Potts_mapping}.
Both representations fulfill the conditions
\begin{align}
  \alpha^3_{A,j} = 1;\ \ \alpha^\dagger_{A,j} = \alpha^2_{A,j} \label{eqn:pfn_props}
\end{align}
for $A = R$ or $L$; moreover, they exhibit `anyonic' commutation relations
\begin{align}\begin{split}
\alpha_{R,j} \alpha_{R,j^\prime} & = e^{i (2\pi/3)\, \text{sgn}(j^\prime-j) } \alpha_{R,j^\prime} \alpha_{R,j} , \\
\alpha_{L,j} \alpha_{L,j^\prime} & = e^{-i (2\pi/3)\, \text{sgn}(j^\prime-j) } \alpha_{L,j^\prime} \alpha_{L,j} \label{eqn:commutation}
\end{split}\end{align}
inherited from the quantum Hall edge fields from which they derive.

Suppose now that the domains in Fig.~\ref{fig:single_trench} are sufficiently narrow that nearest-neighbor parafermionic modes hybridize appreciably.
One can describe the hybridization with a Hamiltonian
\begin{align}\begin{split}
  \tilde{H}_\text{0} &= - \sum_n \Big[ \omega\, ( J\, \alpha^\dagger_{R,2n+1} \alpha_{R,2n} + h\, \alpha^\dagger_{R,2n} \alpha_{R,2n-1} ) \\
  & \qquad\qquad\quad + \text{H.c.} \Big] \label{eqn:single_trench}  \,,
\end{split}\end{align}
where $\omega = e^{i (2\pi/3)}$,
 $n$ sums over unit cells, and the $J$ and $h$ terms couple alternating pairs; see Fig.~\ref{fig:single_trench}.
Physically, these terms reflect tunneling of fractional charges from one domain wall to the next via a path \emph{above} the trench.
(Note that an equivalent form in terms of $\alpha_{L,j}$ is also possible; this form would correspond to tunneling paths below the trench.)

As reviewed in Appendix~\ref{appendix:Potts_mapping} the single-chain Hamiltonian in Eq.~\eqref{eqn:single_trench} maps to the three-state quantum Potts model under a non-local `Fradkin-Kadanoff' transformation akin to the Jordan-Wigner mapping in the Ising model.\cite{Fendley:2012,FradkinKadanoff}
Many properties immediately follow from this identification.
Like its Potts analogue, with $J,h > 0$ the Hamiltonian admits two distinct phases.  For $h > J$ the trench enters a
trivial state where all parafermion modes dimerize and gap out pairwise with their neighbors.  On the other hand, for $J > h$ the parafermion operators dimerize in a shifted pattern leaving `unpaired' modes at the ends of the trench (assuming open boundary conditions).\cite{Fendley:2012}  Here the system exhibits a protected three-fold ground-state degeneracy similar to the topological phase of a Kitaev chain.\cite{Kitaev:2001}
These two phases are separated by a self-dual critical point at $h = J$ described by a non-chiral $\Z3$-parafermion conformal field theory (CFT) with central charge $c = 4/5$.\cite{Dotsenko:3Potts:1984,ZamoFateev:Parafermion:1985}
As our nomenclature hints the right- and left-moving parafermion \emph{fields} at this critical point are closely related to the lattice parafermion \emph{operators} $\alpha_{R/L,j}$.\cite{Mong:2014b}

\begin{figure}
\includegraphics[width=240px]{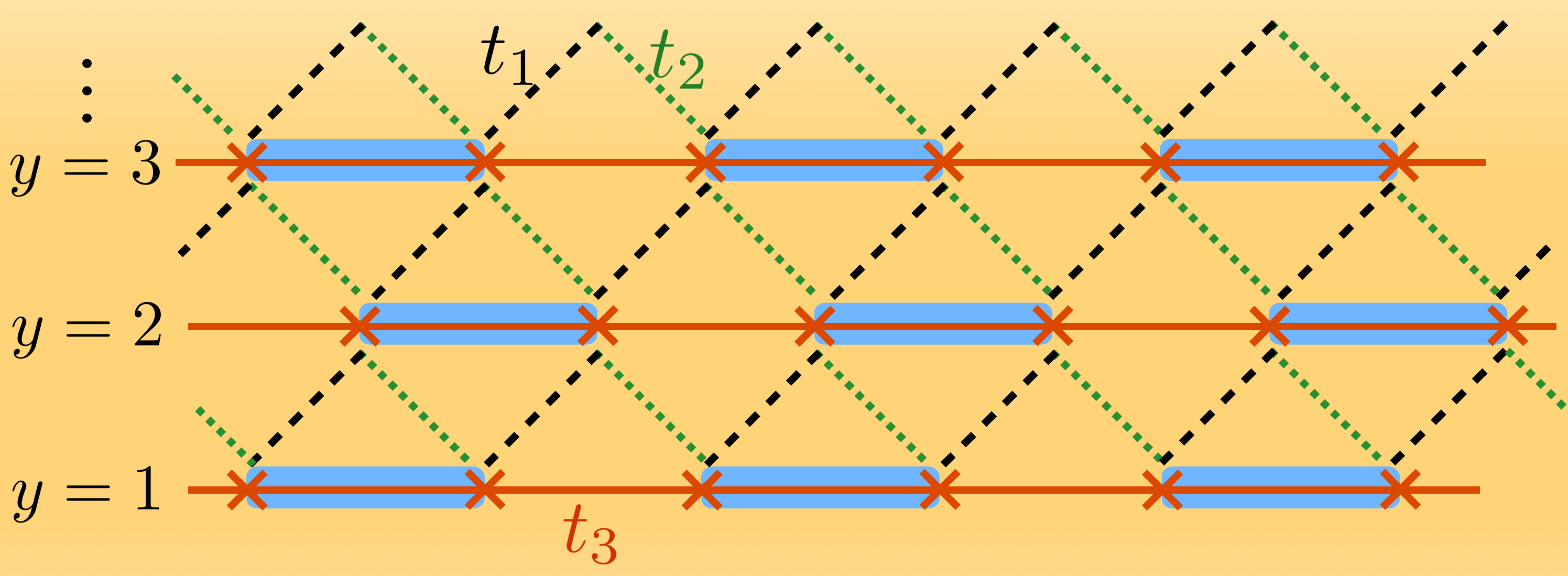}
\caption{A series of one-dimensional parafermion chains interact by tunneling fractional charge through the quantum Hall fluid  [Eq.~\eqref{eqn:interchain}].
For the present work, we consider anisotropic next-neighbor tunneling processes for parafermions arranged in a triangular lattice, allowing us to study (among other limits) the isotropic triangular lattice ($t_1=t_2=t_3$), the isotropic square lattice ($t_1=t_3$, $t_2=0$), and weakly coupled chains ($t_3 \gg t_1,t_2$).
}
\label{fig:2d_trenches}
\end{figure}

Now consider a series of $N$ such trenches at vertical positions $y$, arranged such that parafermion modes reside at the sites of a 2D lattice
(Fig.~\ref{fig:2d_trenches}).
If trenches $y$ and $y+1$ are separated by a finite distance, there will also be fractional-charge tunneling processes \emph{through} the intervening quantum Hall fluid.
Importantly, these processes can only couple the `left' parafermion representation in trench $y$ to the `right' representation in $y+1$.
(The corresponding process with right and left interchanged would pass fractional charge through regions where only whole electrons can pass.)
A general interchain Hamiltonian thus reads
\begin{align}
\tilde{H}_\perp & = - \sum_{y=1}^{N-1} \sum_{j,j^\prime}  \left[  \tilde{t}_{j-j^\prime}\alpha^\dagger_{L,j}(y) \alpha_{R,j^\prime}(y+1) + \text{H.c.} \right] \ .
\label{eqn:interchain}
\end{align}
Because the bulk of each quantum Hall fluid is gapped, the interchain couplings $\tilde{t}_{j-j^\prime}$ should decay rapidly as a function of separation between sites.
Thus taking $\tilde{t}_{j-j^\prime}$ non-zero for only the first- or possibly second-nearest-neighbors in adjacent trenches constitutes a reasonable minimal model.

Figure~\ref{fig:2d_trenches} depicts the specific pattern of interactions that we consider.  Parafermion modes bound to domain walls reside at the sites of
a triangular lattice and couple via anisotropic next-neighbor tunneling processes.
The Hamiltonian of this model reads
\begin{align}\begin{split}
	H & = H_\text{0} + H_\perp \\
	H_\text{0} & = -\sum_{y}  \sum_j \ t_3  \left[\omega\, \alpha^\dagger_{R,j+1}(y) \alpha_{R,j}(y) + \mathrm{H.c.} \right] \\
	H_\perp & = - \sum_{y}  \sum_j \, \Big[ t_1\, \omega\,  \alpha^\dagger_{L,j}(y) \alpha_{R,j}(y+1) \\
		&\qquad + t_2\,  \omega\,  \alpha^\dagger_{L,j}(y) \alpha_{R,j-1}(y+1) + \text{H.c.}\Big]  \:.
	\label{eqn:model}
\end{split}\end{align}
Our intrachain Hamiltonian $H_0$ is identical to Eq.~\eqref{eqn:single_trench} but fixed to \mbox{$J = h \stackrel{\text{\tiny def}}{=} t_3$} such that each horizontal parafermion chain remains tuned to its critical point if decoupled from the other chains.
The interchain bonds then form the triangular lattice pattern of interest, with vertical couplings $t_1$ and slanted couplings $t_2$ in the skewed layout of Fig.~\ref{fig:triangular_lattice}.
Throughout we will assume purely real $t_{1,2,3}$.
Importantly, the Hamiltonian's phase diagram is sensitive to both the magnitude \emph{and} signs for for these couplings, since the only gauge transformations we can make correspond to shifting the parafermion operators by factors of $\omega$.
Notice that the model interpolates between the square lattice ($t_2=0$, $t_1=t_3$), isotropic triangular lattice ($t_1=t_2=t_3$), and decoupled chain ($t_1=t_2=0$) limits---allowing us to explore several important cases in one framework.

Despite the fact that Eq.~\eqref{eqn:model} describes an inherently strongly interacting 2D setup, one can make controlled analytic progress in the weakly-coupled-chain limit with $0 < t_1 = t_2 \ll t_3$.
In this regime the problem reduces to parafermion CFTs in each chain that hybridize to yield a 2D topologically ordered `Fibonacci phase'.\cite{Mong:2014} (See Appendix~\ref{appendix:CFT} for details.)
Quite generally, the Fibonacci phase exhibits the following key properties:

$(i)$ Within the parent quantum Hall fluid, spatial boundaries of the Fibonacci phase support \emph{chiral} $\Z3$ parafermion modes with central charge $c = 4/5$.  These residual gapless edge modes represent `half' of the non-chiral CFT that occurs in each critical chain.

$(ii)$ The bulk supports two quasiparticles types: trivial (fermionic/bosonic) excitations and Fibonacci anyons.

$(iii)$ If the system is defined on a torus or an infinite cylinder, there are two degenerate ground states $\ket{1}$ and $\ket{\varepsilon}$---one associated with each quasiparticle type. \footnote{Here and in point $(iv)$ we specifically intend the labels $\ket{1}$ and $\ket{\varepsilon}$ to
refer to the minimally entangled state (MES) basis of the ground-state subspace.\cite{Zhang:2012q} The MES are the ground states that DMRG generically produces
in calculations on infinitely long cylinders,\cite{Jiang:2012} and which have the specific values of the topological entanglement entropy $\gamma_n$
enumerated in point $(iv)$}

$(iv)$ The entanglement entropy of ground state \mbox{$|n=1,\varepsilon\rangle$} on an infinite cylinder of large but finite circumference $N_y$ scales as $S_n = a N_y - \gamma_n$.
Here \mbox{$\gamma_n = \log(\mathcal{D}/d_n)$} represents the topological entanglement entropy expressed in terms of the anyon quantum dimensions $d_1 = 1$, $d_\varepsilon = \varphi$ and the total quantum dimension $\mathcal{D}=\sqrt{1+\varphi^2}$, with $\varphi = (1+\sqrt{5})/2$ the golden ratio.

$(v)$ The entanglement spectrum\cite{Li:2008e}---which can be computed from the CFT\cite{QiKatsuraLudwig}---resembles the energy spectrum of the physical edge with chiral central charge $4/5$.
The entanglement spectrum of $\ket{1}$ consists of states descended from primaries $\{1,\psi,\psi^\dag\}$ (with scaling dimensions $\{0,\frac23,\frac23\}$ respectively),
while the spectrum of $\ket{\varepsilon}$ consists of descendents of $\{\varepsilon,\sigma,\sigma^\dag\}$ (with scaling dimensions $\{\frac25,\frac1{15},\frac1{15}\}$ respectively).
The specific counting\cite{Henkel} of states appears in Fig.~\ref{fig:Ny4_tperp_es}.

Outside of the highly anisotropic, weakly-coupled-chain limit, the analytical methods used to establish the preceding results break down entirely.
Addressing the broader phase diagram of the model---particularly the extent of the Fibonacci phase---is the central goal of this paper.
For this we turn to density matrix renormalization group (DMRG) calculations of the ground states of Eq.~\eqref{eqn:model}, primarily on infinitely long cylinders
using the infinite DMRG algorithm proposed in Ref.~\onlinecite{McCulloch:2008}.
To implement the Hamiltonian in Eq.~\eqref{eqn:model} we do not work directly with parafermion operators [Eqs.~\eqref{eqn:pfn_props}--\eqref{eqn:commutation}].
Instead, following Appendix~\ref{appendix:dmrg_path} we map the parafermion degrees of freedom to $\Z3$ clock variables, which provides a much more convenient (but formally equivalent) representation for numerics.

It is worth noting some technical features of the model in Eq.~\eqref{eqn:model} that facilitate our DMRG studies.
Retaining all possible first- and second-neighbor interchain couplings in a square-lattice arrangement of the parafermion sites
would lead to a one-dimensional Hamiltonian (as seen by the DMRG algorithm) with interactions up to a range $d_\text{max} = 2 N_y$,
where $N_y$ is the circumference of the quasi-2D cylinder used for the simulation.
In contrast, the pattern of triangular lattice couplings we study here (Fig.~\ref{fig:triangular_lattice}) gives a one-dimensional Hamiltonian with a maximum range
scaling only as $d_\text{max} = N_y$.
Additionally, as we discuss in the next section, the two-chain limit $(N_y = 2)$ of the triangular model with open $y$-boundary conditions
is self-dual for a certain choice of parameters.
Appealing to self-duality allows us to exactly determine the location of an important line of critical points that roughly represents a remnant of the Fibonacci phase compressed into a two-chain system.
Finally, the properties of the bona fide 2D Fibonacci phase enumerated above provide sharp numerical fingerprints that we can use to track the phase diagram in our multi-chain simulations $(N_y>2)$, which we describe in Sec.~\ref{sec:cylinders}.

\section{Two-Chain Limit \label{sec:two_chain}}

\begin{figure}[b]
\includegraphics[width=180px]{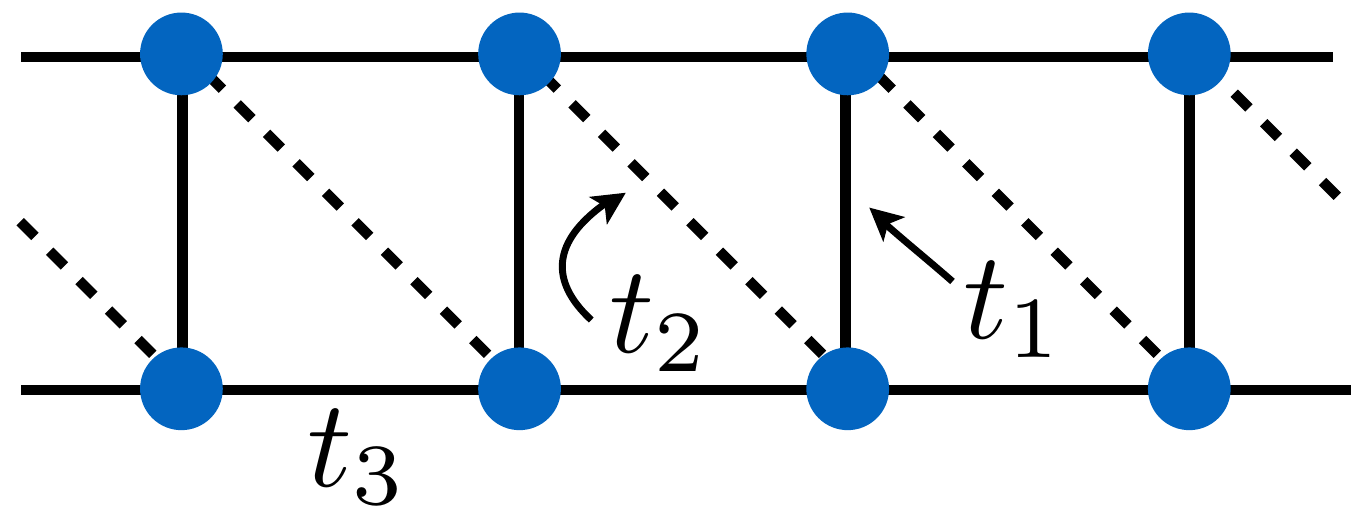}
\caption{Two-chain limit of the Hamiltonian in Eq.~\eqref{eqn:model}, with open boundary conditions along the $y$ direction.}
\label{fig:two_chain_model}
\end{figure}

Before considering the model Eq.~\eqref{eqn:model} on multi-leg cylinders, it is helpful to understand the limit of only two coupled chains
with open boundary conditions in the $y$ direction (see Fig.~\ref{fig:two_chain_model}).
As we will see this limit already contains precursors of the 2D Fibonacci phase that we will uncover later on when studying wider geometries.
Throughout this section we assume $t_3 \geq 0$ for simplicity but consider either sign of $t_{1,2}$.

\begin{figure}
\includegraphics[width=120px]{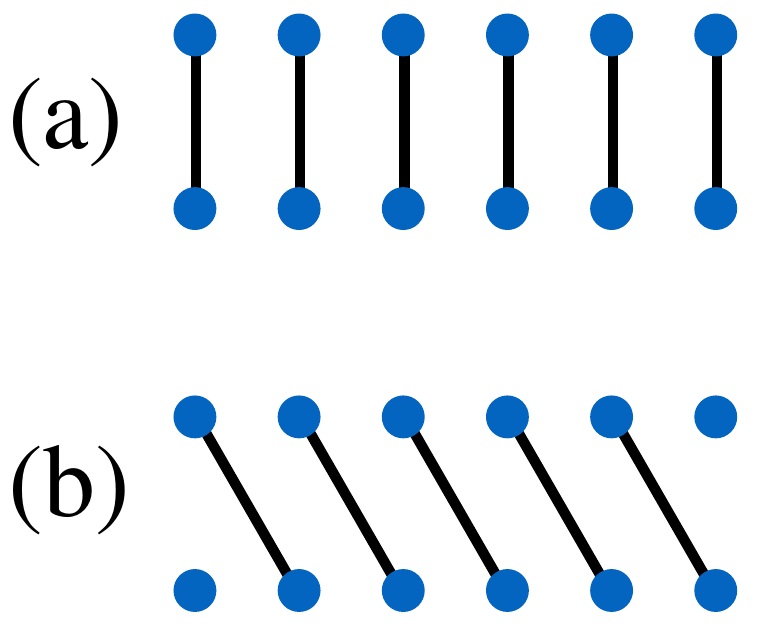}
\caption{Pairing of sites in the ground state for the two-chain ladder system in the limits of (a) dominant $t_1>0$, which produces a trivial gapped phase, and (b) dominant $t_2 > 0$, which produces a topological gapped phase with a three-fold ground-state degeneracy.}
\label{fig:ladder_pairing}
\end{figure}

To understand the  phase diagram of the two-chain system, first consider the limit where $t_1>0$ greatly exceeds both $t_2$ and $t_3$.
In the extreme case with $t_2 = t_3 = 0$, the ground state is found by pairing each parafermion site with the one directly above or below it, yielding the trivial product state
illustrated in Fig.~\ref{fig:ladder_pairing}(a).
(Saying that two parafermions $\alpha_i$ and $\alpha_j$  `pair' means that they form an eigenstate of $\frac{1}{2} (\omega\, \alpha^\dagger_i \alpha_j + \text{H.c.})$ with maximal eigenvalue 1.)
Because this limit supports a unique ground state protected by a robust gap, this trivial phase persists upon restoring sufficiently small $t_{2,3}$ couplings.

Consider next the limit with $t_2>0$ much larger than $t_1$ and $t_3$.
With $t_1 = t_3 = 0$ the ground state again arises by pairing sites, but now in the skewed pattern depicted in Fig.~\ref{fig:ladder_pairing}(b).
Any finite ladder with open boundaries along the horizontal direction thus contains one unpaired site at each end---signifying a topologically nontrivial phase.
These two decoupled end sites, taken together, form a degenerate three-level system.
It follows that there are three degenerate ground states distinguished by their `triality', defined as
\begin{align}
Q = \prod_n \omega\alpha^\dagger_{2n} \alpha_{2n-1} \ ,
\end{align}
which admits eigenvalues 1, \mbox{$\omega$}, or \mbox{$\omega^2$}.
Restoring small, finite $t_1$ and $t_3$ only splits the ground-state degeneracy by
an exponentially small amount in the system size, as the processes mixing the end states require tunneling
across a macroscopic number of sites. (Note that this argument applies only to the ground states. The excited states, strictly three-fold degenerate for \mbox{$t_1=t_3=0$},
generally split by an amount decaying only as a power law in the system size.\cite{Jermyn})

Intuitively, one expects a 1D phase transition between the gapped states sketched in Fig.~\ref{fig:ladder_pairing} when $t_1$ and $t_2$ become comparable.
To be more quantitative, we invoke a weakly-coupled-chain analysis and find an interesting scenario.
In the strictly decoupled-chain limit ($t_{1,2}=0$), the low-energy properties of each chain can be described by a $\Z3$ parafermion CFT at $c=4/5$ with a pair of counter-propagating chiral fields.\cite{ZamoFateev:Parafermion:1985}
Turning on weak interchain couplings $t_{1,2}$ generically gaps all of these fields, but when one fine-tunes $t_1 = t_2 >0$ the
top chain's left-mover and the bottom chain's right-mover remain gapless.\cite{Mong:2014}
Thus along the line $t_1=t_2$ (at least for $0 < t_{1}, t_{2} <\!\!< t_3$) the two-chain system should be critical and described 
by a \emph{single} non-chiral $\Z3$ parafermion CFT in which the right- and left-movers have, in a sense, spatially separated along the vertical direction.
We can, however, actually reach a much stronger conclusion.
Appendix~\ref{appendix:ladder_duality} shows that for $t_1 = t_2$ the two-chain Hamiltonian maps to itself under duality followed by a time-reversal transformation.
Hence there should exist a continuous phase transition line \emph{precisely} along $t_1 = t_2>0$ for \emph{any} $t_3>0$.

\begin{figure}
\includegraphics[width=\columnwidth]{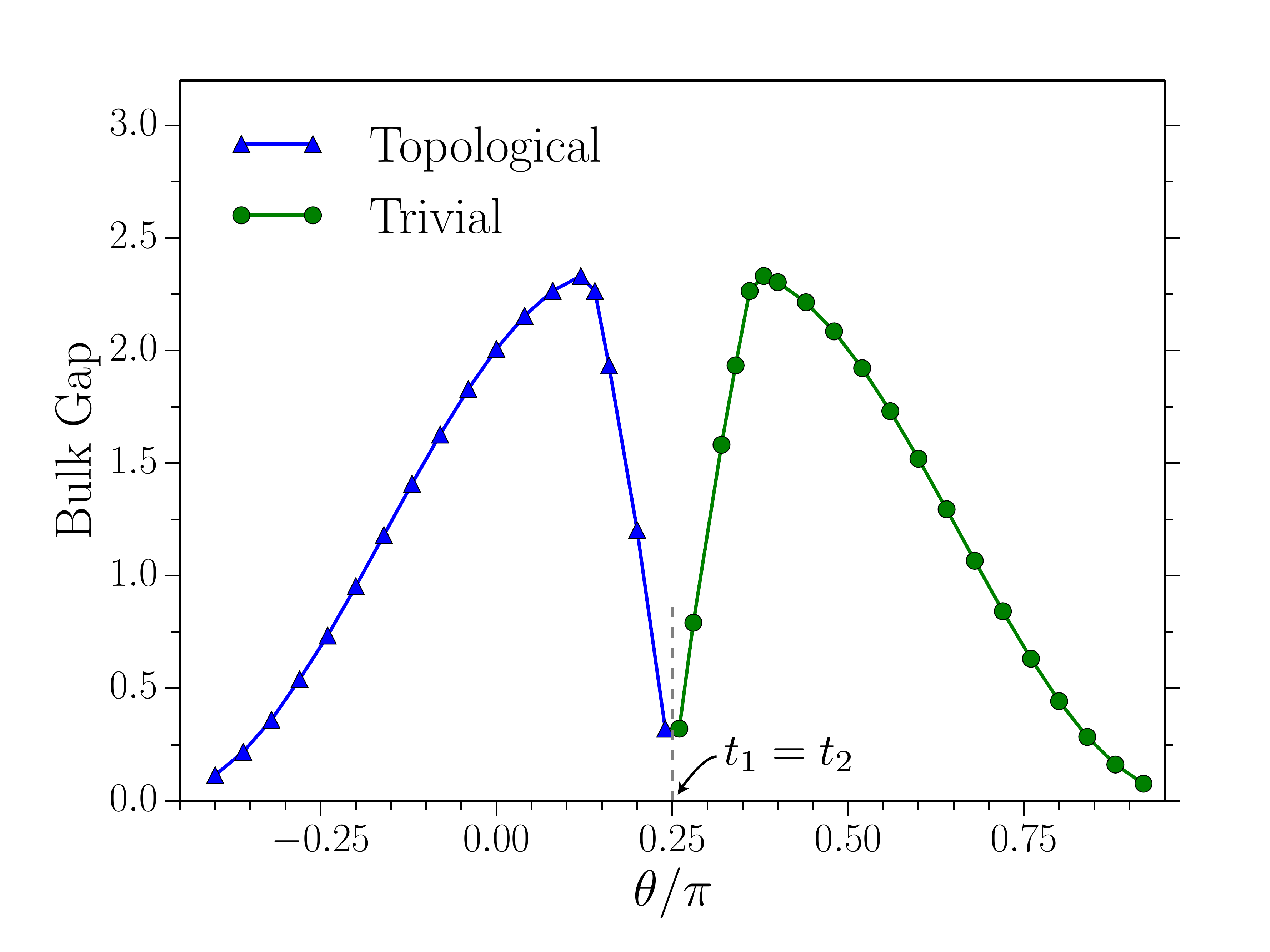}
\caption{Bulk gap of the two-chain system (Fig.~\ref{fig:two_chain_model}) as a function of $\theta$ with \mbox{$t_1 =  \sin\theta$}, \mbox{$t_2= \cos\theta$},
and $t_3=1$.
The vertical dashed line marks the \mbox{$t_1=t_2\,(=1/\sqrt{2})$} critical point separating topological and trivial gapped phases. For the topological phase, the first excited state is in the quantum number sector
with triality 1; for the trivial phase, the first excited states are two-fold degenerate and lie in the $\omega$ or $\omega^2$ triality sectors.}
\label{fig:r1.0_gaps}
\end{figure}

We confirm all of the above predictions---and extend them beyond the analytically tractable regimes---using DMRG.
Unless otherwise stated we use the infinite DMRG algorithm\cite{McCulloch:2008} to reach
 the thermodynamic limit in the horizontal direction while taking open boundary conditions along the vertical direction.
For both gapped phases we find essentially exact results (truncation errors below $10^{-12}$)
by keeping only a few hundred states in DMRG.
First, we check for the presence of the critical line at $t_1=t_2$ by computing bulk gaps using infinite boundary conditions.\cite{Phien:2012}
As shown in  Fig.~\ref{fig:r1.0_gaps} the gap indeed closes precisely at $t_1=t_2$.
(In the horizontal axis $\theta$ parameterizes the couplings through $t_1=  \sin\theta$  and $t_2= \cos\theta$ with $t_3=1$.)

\begin{figure}
\includegraphics[width=\columnwidth]{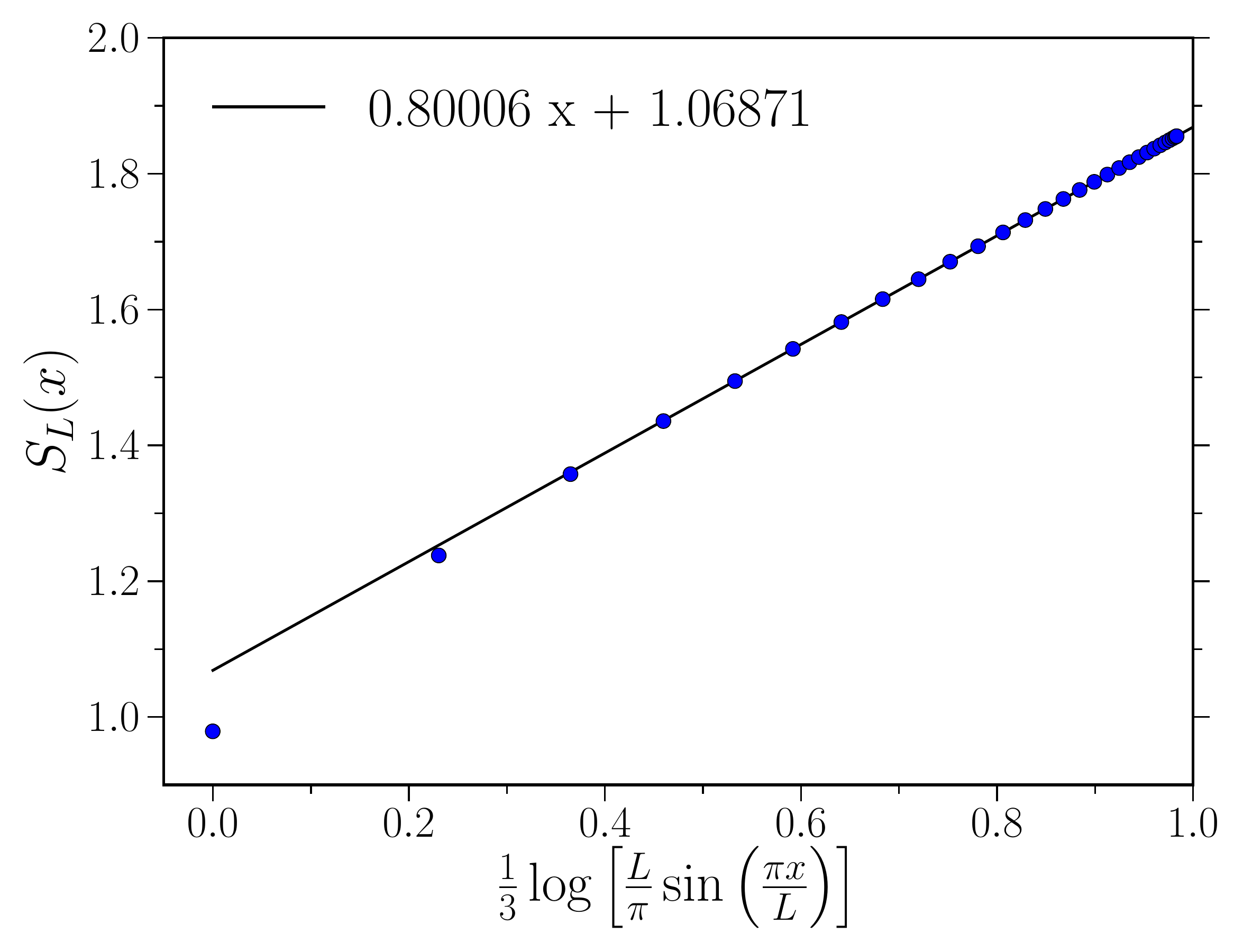}
\caption{Entanglement entropy $S_L(x)$ for a length-$x$ subregion of a periodic
two-chain system with length $L=60$.  Data were obtained at the critical point using parameters $t_1/t_3 = t_2/t_3 = 1$.
By fitting to the CFT prediction we extract a central charge $c = 4/5$, consistent with analytical arguments.
}
\label{fig:central_charge}
\end{figure}

To additionally extract the central charge along this critical line, we use DMRG to study periodic systems of size $L$ along the horizontal direction and
compute the entanglement entropy $S_L(x)$ of a subregion of size $x$.
For a CFT with central charge $c$ the entanglement is predicted to scale as\cite{Calabrese:2009}
\begin{align}
S_L(x) = \frac{c}{3} \ln\Big[  \frac{L}{\pi} \sin\Big(\frac{\pi x}{L}\Big) \Big] \:.
\end{align}
Figure~\ref{fig:central_charge} displays our simulation results.  For $t_1/t_3=t_2/t_3=1$ we find $c=4/5$ to very high accuracy,
confirming the prediction that, along the critical $t_1=t_2$ line, the coupled chains host a \emph{single} pair of gapless $\Z3$
parafermion CFT edge modes.

\begin{figure}
\includegraphics[width=\columnwidth]{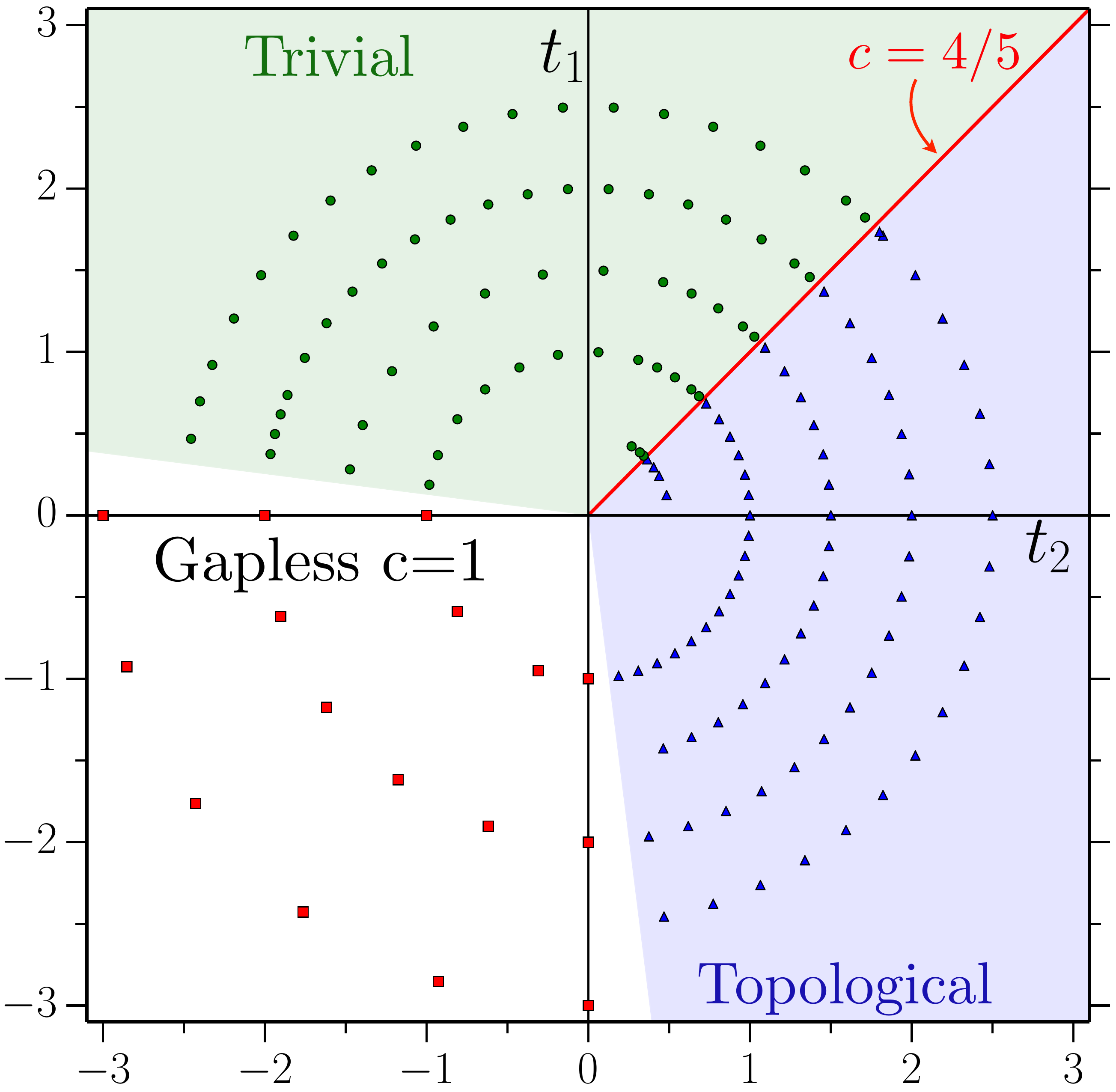}
\caption{Full phase diagram of the two-chain system in units where $t_3=1$. The trivial phase (green circles) surrounds the $t_1 > 0$ axis while the topological phase (blue triangles) surrounds the $t_2 > 0$ axis.  The remainder of the phase diagram shown is gapless with central charge $c=1$, which we verified numerically
for points indicated with red squares.  Transition lines separating the gapless state from the two gapped phases are approximate.  The critical point along the line $t_1 = t_2$ is exact, however, and represents a precursor to the Fibonacci phase appearing in the 2D limit.
}
\label{fig:2leg_phase_diagram}
\end{figure}

Figure~\ref{fig:2leg_phase_diagram} shows the full $t_3\geq 0$ phase diagram of the two-chain system computed with DMRG.
We numerically identify the trivial gapped phase occurring in the large $t_1 > 0$ region by
 searching for the presence of a finite gap and a unique ground state.
To identify the region lying in the topological phase stabilized for large $t_2 > 0$,
we primarily examine the entanglement spectrum within DMRG. For the entire topological phase each entanglement ``energy'' exhibits a robust
three-fold degeneracy. At various points within this phase we also checked that finite systems with open boundary conditions possess three degenerate ground states.

In addition to the above two gapped phases, we also find a large region for either $t_1$ or $t_2$ negative where the system is gapless.
Numerically computing the central charge using the same approach as for Fig.~\ref{fig:central_charge}, we find this entire gapless region has central charge \mbox{$c = 1$}.
To understand this phase, consider the limit \mbox{$t_1 < 0$} and $t_2,t_3\rightarrow 0$. In this limit the antiferromagnetic $t_1$ coupling favors two of the three states on each rung, effectively projecting out the third. This allows a natural mapping to a spin $1/2$ chain.
Perturbatively reintroducing $t_2$ and $t_3$ gives a Hamiltonian of XXZ type. Close to the negative $t_1$ axis, the effective Z coupling is
parametrically suppressed relative to the XY coupling, leading to a gapless XY phase that is known to have $c=1$.

For our purposes the most interesting feature of the two-chain phase diagram is the $c = 4/5$ transition line occurring at $t_1 = t_2$ where, remarkably,
the low-energy right- and left-movers residing at opposite sides of the ladder do not hybridize even for large $t_{1,2}$.
This critical point reflects a maximally `squeezed' cousin of the stable 2D Fibonacci phase; in the latter, the right- and left-movers are separated by \emph{macroscopic} distances and therefore coexist harmoniously without any fine-tuning.
In the next section we will show that upon adding further chains this line does in fact open into an extended region that develops into the Fibonacci phase in the 2D limit.

\section{Towards two dimensions}\label{sec:cylinders}

Armed with our insights from the two-chain model explored in the previous section, we wish to now deduce the phase diagram for our parafermion model in the full 2D limit.
For this purpose we use extensive DMRG simulations to complement earlier analytical studies that apply only in the case of weakly coupled chains ($t_3 \gg t_{1,2}$).
We are particularly interested in tracking the extent of the Fibonacci phase as the interchain coupling increases away from this limit, though we will attempt to address the nature of nearby states as well.
To approach two dimensions we study the Hamiltonian of Eq.~\eqref{eqn:model} on cylinders with $N_y = 4, 6, 8$, and 10 sites around the circumference.
The cylinders will be taken infinitely long in the horizontal direction, in part to avoid complications arising from possible gapless edge states.
For our calculations we typically retain up to $m=5500$ states in DMRG for truncation
errors up to $10^{-8}$ and often as small as $10^{-10}$.

\subsection{Fibonacci Phase on an Anisotropic Triangular Lattice}

Until specified otherwise we assume $t_{1,2,3} \geq 0$ and fix $t_1 = t_2 \stackrel{\text{\tiny def}}{=} t_\perp$ (Secs.~\ref{sec:fib_phase_diagram} and \ref{sec:2d_phase_diagram} relax these assumptions).
Our goal here is to numerically assess how the system evolves as we vary $t_\perp/t_3$ to tune from the decoupled-chain limit $t_\perp=0$ up through the isotropic triangular lattice point $t_\perp=t_3$ and beyond.
From the coupled-wire analysis, one expects the 2D system to immediately enter the gapped Fibonacci phase upon turning on any small but finite $t_\perp/t_3 \ll 1$.
In this limit the system should exhibit a quite long correlation length in the horizontal direction but---crucially---an arbitrarily short correlation length along $y$.
This feature is extremely attractive for DMRG: it implies that infinite cylinders even with relatively small $N_y$ possess local properties closely emulating those of the fully 2D system of interest.
Our simulation results, even for cylinders as small as $N_y=4$, indeed show strong evidence that the system realizes the Fibonacci
phase once $t_\perp > 0$.
In what follows we will numerically recover the characteristics $(i)$ through $(v)$ of the Fibonacci phase delineated near the end of Sec.~\ref{sec:models}.

\begin{figure}[t]
\includegraphics[width=\columnwidth]{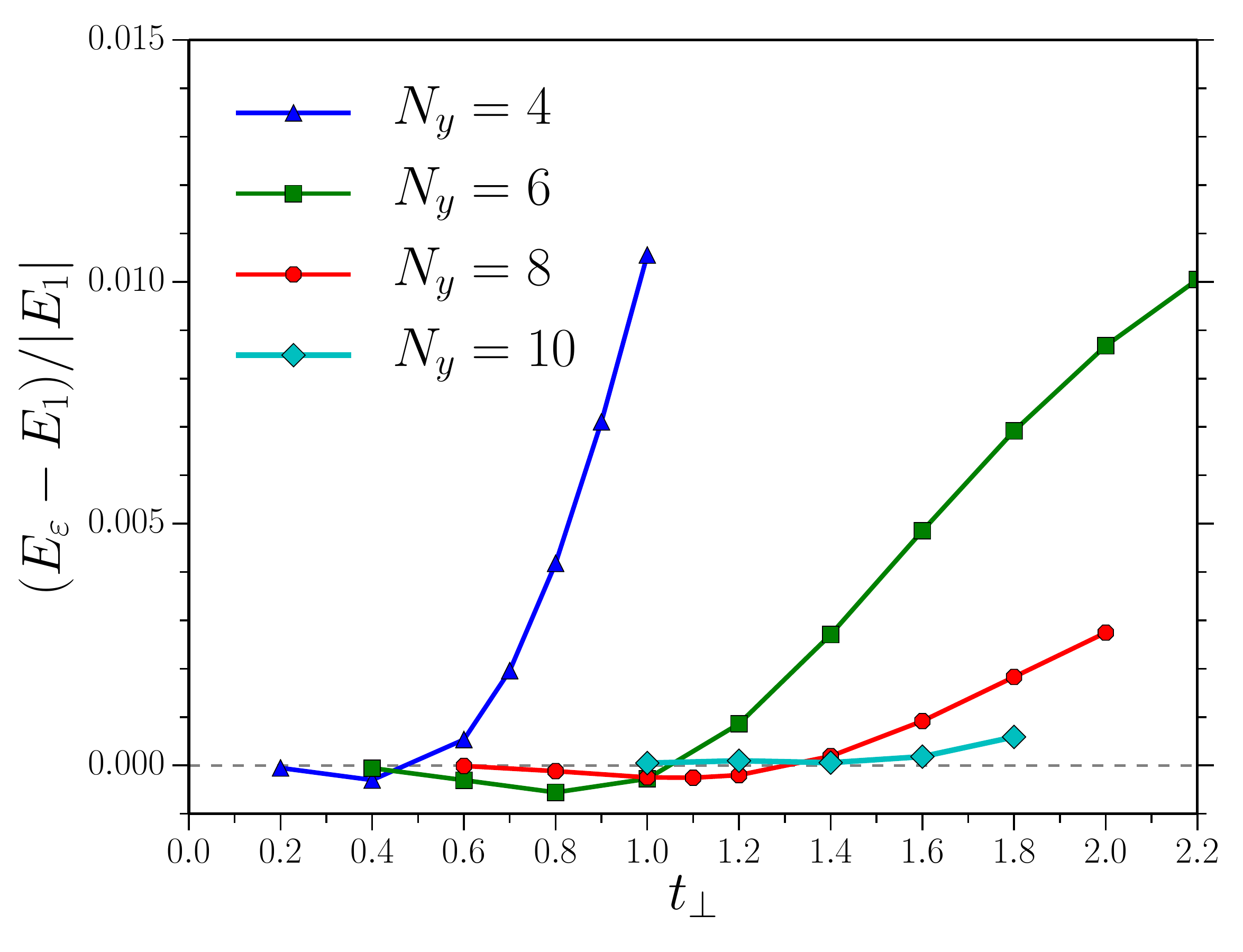}
\caption{ Relative energy splitting between quasi-degenerate ground states of cylinders with $N_y=4$--$10$  as a function of $t_\perp$ (in units where $t_3=1$).
Notice that for $N_y \geq 8$  the states remain nearly degenerate even up to the isotropic triangular lattice point $t_\perp = 1$.
This is the first evidence suggesting that in the 2D limit the Fibonacci phase enjoys a wide stability window extending from weakly coupled chains past the isotropic triangular lattice limit.}
\label{fig:tperp_endiff}
\end{figure}

For a wide range of $t_\perp/t_3$ we observe two quasi-degenerate ground states---both in the triality 1 quantum number sector---obtained by starting the infinite DMRG algorithm in randomized initial states.
(Which quasi-degenerate ground state appears depends on the precise initial state used.)
As we will argue later based on entanglement-entropy scaling, these two states span the entire ground-state subspace.
Anticipating Fibonacci-phase physics, let us call the quasi-ground state with lower entanglement entropy $\ket{1}$ and the other  $\ket{\varepsilon}$.
Figure~\ref{fig:tperp_endiff} shows the relative energy splitting  \mbox{$(E_\varepsilon-E_1)/|E_1|$} of these states as a function of $t_\perp/t_3$, at various system sizes.
For a fixed circumference $N_y$ and sufficiently weak $t_\perp/t_3$, we observe a very small splitting ($< 0.1\%$) with weak $t_\perp$ dependence.
Beyond an $N_y$-dependent scale of $t_\perp$, however, the splitting grows rapidly with the interchain coupling.
We expect that the crossover between these behaviors transpires when the correlation length in the $y$ direction becomes comparable to the circumference; this interpretation is consistent with the enhanced robustness of the degeneracy evident in Fig.~\ref{fig:tperp_endiff} upon increasing $N_y$.

\begin{figure}
\includegraphics[width=\columnwidth]{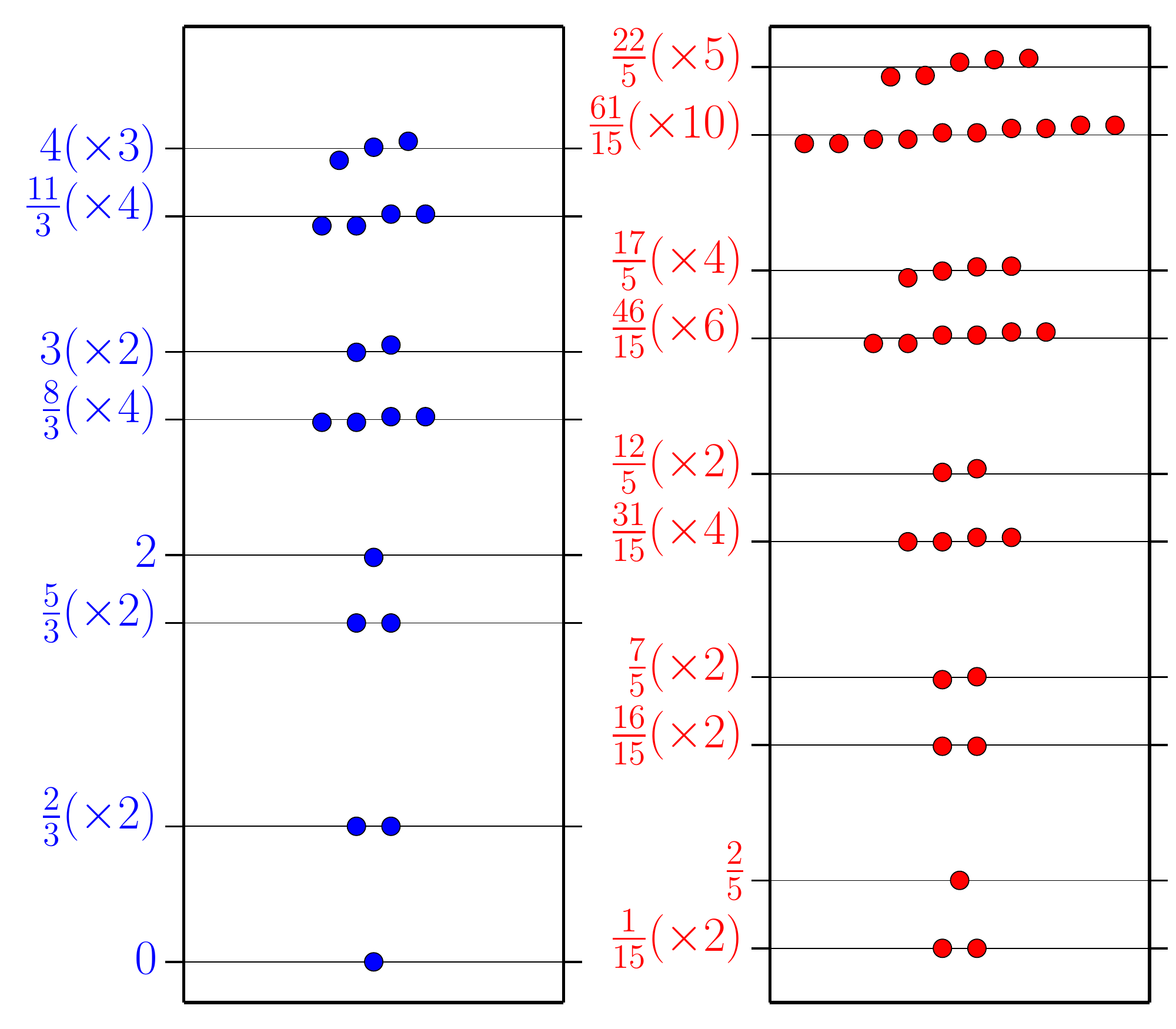}\\
	\hspace{12mm}\mbox{\large $\ket{1}$}\hspace{40mm}\mbox{\large $\ket{\varepsilon}$}
\caption{Entanglement ``energies'' of the quasi-ground states $\ket{1}$ (left column) and $\ket{\varepsilon}$ (right column) for the $N_y=4$,
\mbox{$t_\perp/t_3=0.2$} cylinder after shifting and rescaling the spectra
to match the lowest two levels of the $\Z3$ parafermion CFT prediction. The numbers in parentheses indicate the predicted degeneracies.
The same rescaling was used for both spectra, but with different shifts; thus the fit requires only three fitting parameters. }
\label{fig:Ny4_tperp_es}
\end{figure}

Next we examine the entanglement spectrum of both quasi-ground states, which for a topological phase is expected to reveal the gapless
low-energy spectrum in the presence of an open edge.\cite{Li:2008e} Following the approach of Ref.~\onlinecite{Cincio:2013}, we plot the entanglement
spectrum after applying a non-universal overall shift and rescaling such that the lowest two levels match the spectrum of the chiral $\Z3$ parafermion CFT.
(Each entanglement ``energy'' $\epsilon_i$ is defined in terms of a reduced-density-matrix eigenvalue $p_i$ through \mbox{$\epsilon_i = -\ln{p_i}$}.)
As shown in Fig.~\ref{fig:Ny4_tperp_es} for $N_y=4$ and $t_\perp/t_3=0.2$, all the remaining entanglement energies
exhibit the same pattern of degeneracies and relative energy-level spacings as the excitations of the chiral $\Z3$ parafermion CFT on a ring.
Specifically, we find that the entanglement spectrum of state $\ket{1}$ matches the CFT level pattern for the superselection sector corresponding to the primary
fields $1, \psi, \psi^\dagger$ (and their descendants), while the state $\ket{\varepsilon}$ corresponds to the fields $\varepsilon,\sigma, \sigma^\dagger$.
From the fusion algebra of the chiral fields, this suggests that $\ket{\varepsilon}$ and $\ket{1}$ respectively carry Fibonacci and identity flux.
We observed similar agreement of the entanglement spectrum with the CFT prediction for other (small enough) values of $t_\perp/t_3$ and for $N_y=6,8,10$,
but choose to show the $N_y=4$ results to emphasize how quickly the 2D behavior sets in for these anisotropic cylinders.

For further evidence that we are observing a topologically ordered phase, we compute the topological entanglement entropy $\gamma_n$ for quasi-ground states $|n = 1,\varepsilon\rangle$.
For a ``vertical'' entanglement cut dividing our infinite cylinders into two semi-infinite halves, the entanglement entropy is predicted to scale as\cite{Kitaev:2006t,Levin:2006}
\begin{equation}
S_n = a N_y - \gamma_n + \cdots
	\label{eq:TEE}
\end{equation}
where the first term encodes the boundary law expected to hold within a gapped phase. 
Because topologically ordered states are locally indistiguishable in the 2D limit, the coefficient $a$ should be independent of the ground state $n$
(although it is sensitive to other microscopic details).
Furthermore, because DMRG is somewhat biased to low-entanglement states, it generally favors
ground states of topological phases on infinite cylinders having well-defined topological flux through the cylinder, known as minimum entropy
states.\cite{Zhang:2012q}
For such states, $\gamma_n = \log(\mathcal{D}/d_n)$, with $d_n$ the quantum dimension of the anyon type associated with the $n$\textsuperscript{th} ground state
and $\mathcal{D} = \sqrt{\sum_n d^2_n}$ is the total quantum dimension of the theory.
Notice that $\sum_n e^{-2\gamma_n} = 1$ when summed over~$n$, providing a way to check whether a complete set of ground states has been found.

\begin{figure}[t]
\includegraphics[width=\columnwidth]{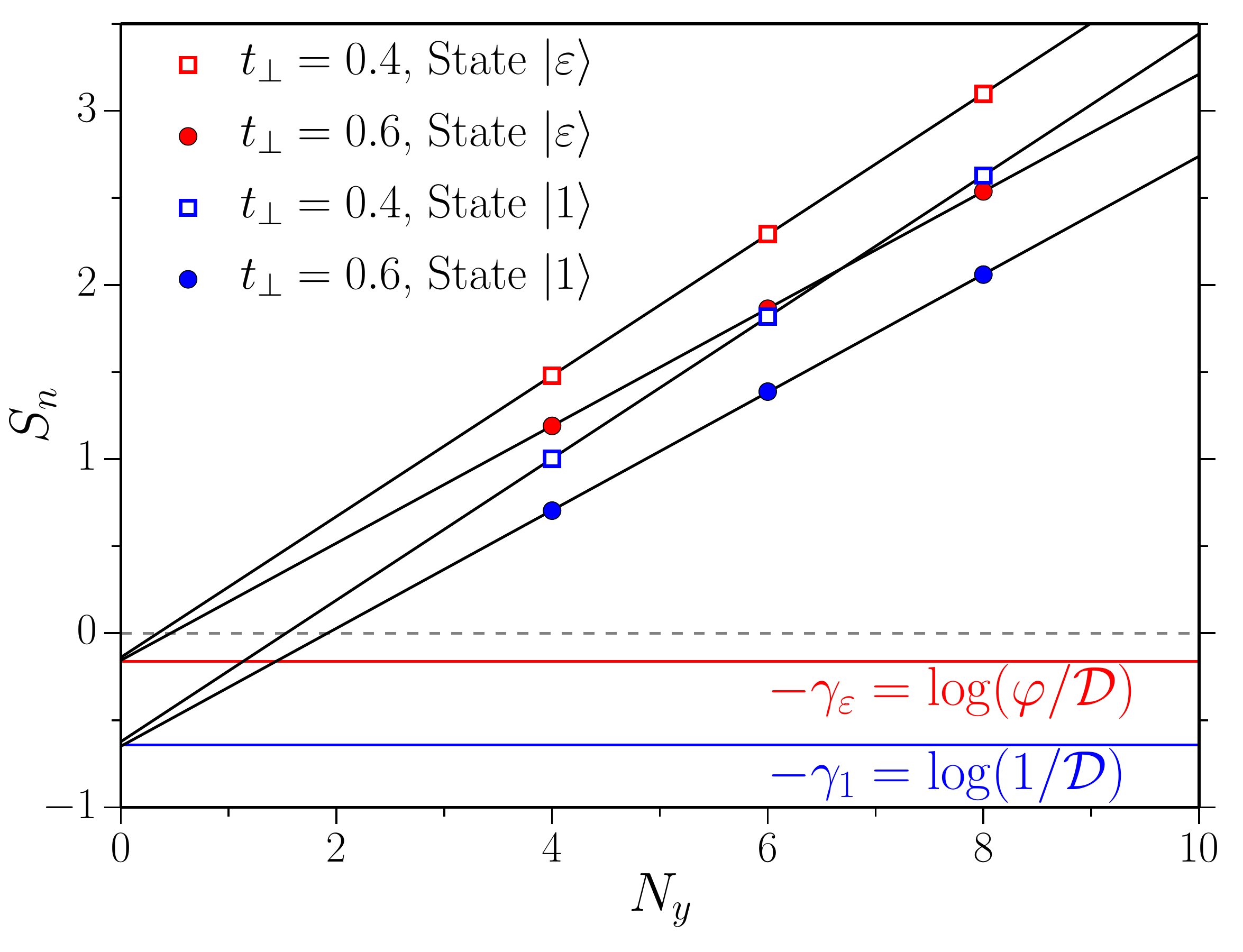}
\caption{Entanglement entropy fits for $N_y=4,6,8$ cylinders with two different interchain couplings $t_\perp$ (with $t_3=1$).
All entanglement-entropy
values are shifted downward by $\log\sqrt{3}$ as explained in the text.
The $y$-intercepts closely match the topological entanglement entropies $\gamma_n$ predicted for the Fibonacci phase.}
\label{fig:S_fits}
\end{figure}

Figure~\ref{fig:S_fits} fits our numerically computed entanglement entropies to the form in Eq.~\eqref{eq:TEE}.
One subtlety here arises from the fact that the parafermions in our system can not appear in vacuum, but rather require a host system---i.e., a $\nu=2/3$ quantum Hall state or similar `parent' topological phase.
Consequently, to back out the $\gamma_n$'s of interest we must shift our entanglement measurements by the topological entanglement entropy of the parent phase, $-\gamma_{(2/3)} = -\log\sqrt{3}$.
(The unshifted $\gamma_n$ values turn out to be negative.)
Table~\ref{table:tee} shows that after applying the shift, the observed $\gamma_n$'s for two different magnitudes of $t_\perp$
agree very well with the theoretical prediction for the Fibonacci phase quoted in Sec.~\ref{sec:models}: $d_1 = 1$, $d_\varepsilon = \varphi$ and $\mathcal{D} = \sqrt{1+\varphi^2}$,
where $\varphi = (1+\sqrt{5})/2$ is the golden ratio.
Furthermore, $(e^{-2\gamma_1} + e^{-2\gamma_\varepsilon})$ is very close to unity for both $t_\perp$ values, allowing us to deduce
that the system admits no further ground states beyond $\ket{1}$ and $\ket{\varepsilon}$.

\begin{table}[t]
	\renewcommand{\arraystretch}{1.1}
	\begin{tabular}{ l | c | c | c }
		& $\gamma_1$ & $\gamma_\varepsilon$ & $\displaystyle e^{-2 \gamma_1}+e^{-2 \gamma_\varepsilon}$  \\
	\hline\hline
	Exact &  $\log\mathcal{D} \approx 0.6430$ & $\log(\mathcal{D}/\varphi) \approx 0.1617$ & 1  \\
	\hline
	$t_\perp = 0.4$\textsuperscript{a} & 0.6235 & 0.1393 & 1.0442 \\
	$t_\perp = 0.4$\textsuperscript{b} & 0.6306 & 0.1538 & 1.0186 \\
	$t_\perp = 0.6$ & 0.6498 & 0.1562 & 1.0043 \\
	\end{tabular}
	\caption{Intercept values extracted from the fits in Fig.~\ref{fig:S_fits} for $t_\perp=0.4, 0.6$ and $t_3 = 1$.
		Both agree with the theoretically predicted topological entanglement entropy shown in the table's first row, and demonstrate that the set of ground states is complete.  (The latter conclusion follows from the fact that the data in the right column are very close to unity.)
		For the $t_\perp=0.4$ system we give both (a) the intercepts shown in Fig.~\ref{fig:S_fits} obtained from fitting all three $N_y=4,6,8$ points and (b) intercepts obtained from only fitting the $N_y=4,6$ entropies that could be computed more accurately.}
	\label{table:tee}
\end{table}

\begin{figure}[t]
\includegraphics[width=\columnwidth]{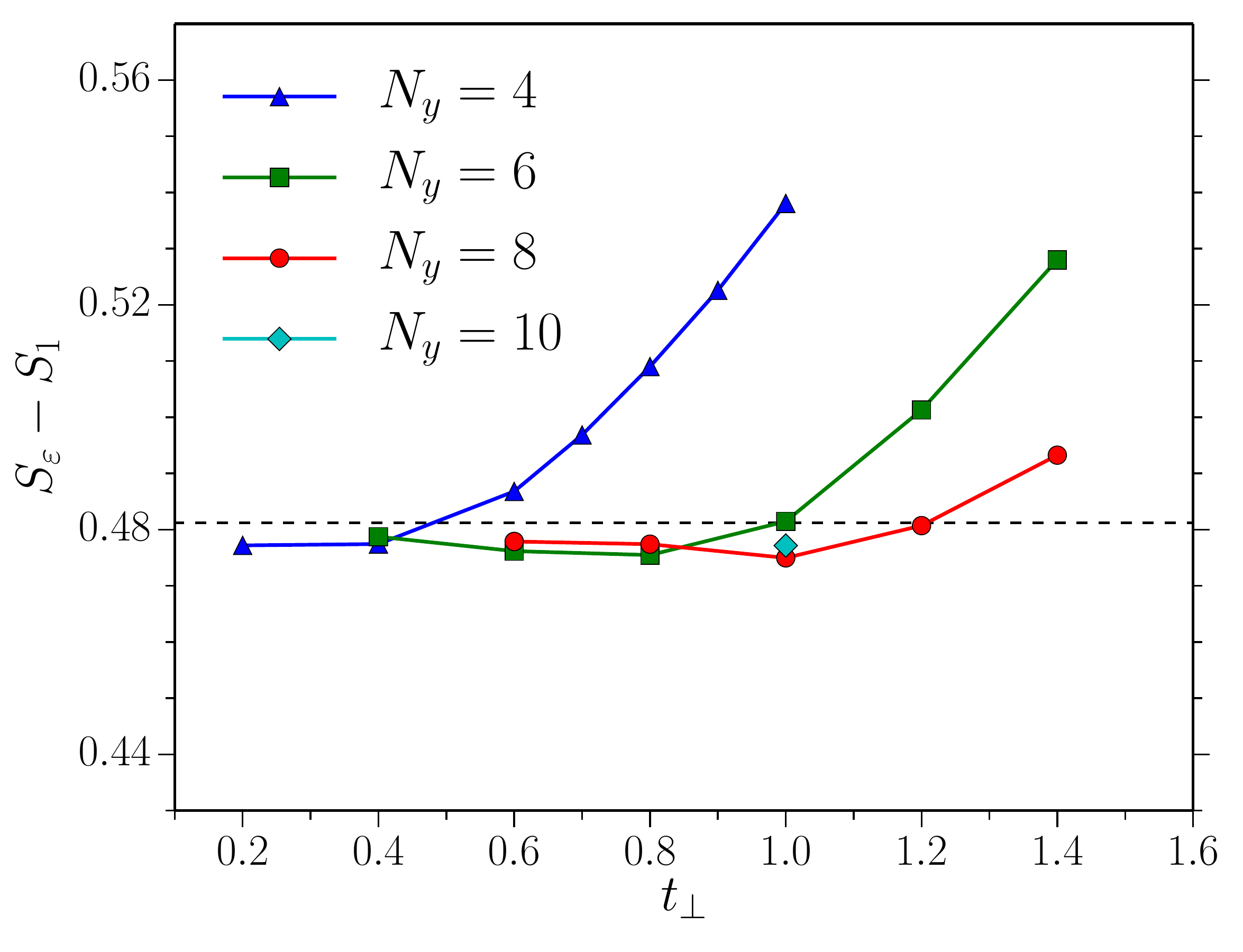}
\caption{Entanglement entropy difference between the quasi-degenerate ground states of $N_y=4,6,8, 10$ cylinders as a function of interchain coupling $t_\perp$ (with $t_3=1$).
              The horizontal dashed line denotes the thermodynamic-limit prediction $\log\varphi \approx 0.481$.
              Our $N_y = 8$ and 10 data further indicate that the Fibonacci phase survives even beyond the isotropic triangular lattice point in the 2D limit, corroborating
              the evidence presented in Fig.~\ref{fig:tperp_endiff}.
}
\label{fig:Sdiff}
\end{figure}

The difference \mbox{$S_\varepsilon-S_1$} of entanglement entropies for the ground states has been found to converge more rapidly as a function of $N_y$ than  linear fits of each individual $S_n$.\cite{Cincio:2013}
Within the Fibonacci phase, we expect
\begin{equation}
S_\varepsilon-S_1 = -\gamma_\varepsilon + \gamma_1 = \log\varphi \approx 0.481
\end{equation}
provided the cylinder size exceeds the correlation length.
The numerical results for $S_\varepsilon-S_1$ in Fig.~\ref{fig:Sdiff} show good agreement with this value, especially for larger and/or more anistropic cylinders which are expected to have the weakest finite-size effects.

Our numerical evidence regarding the entanglement entropy, entanglement spectra, and ground-state degeneracy together strongly indicate the onset of a Fibonacci phase over a wide range of parameters.
	\footnote{The data are also consistent with the time-reversal of the Fibonacci phase.}
These results not only corroborate analytical findings for the strongly anisotropic limit with $t_\perp/t_3 \ll 1$; quite remarkably, Figs.~\ref{fig:tperp_endiff} and \ref{fig:Sdiff} also reveal that the Fibonacci phase persists into the isotropic-triangular-lattice case \mbox{$t_\perp/t_3 = 1$} and beyond!

\subsection{Extent of the Fibonacci Phase}
\label{sec:fib_phase_diagram}

\begin{figure}[t]
\includegraphics[width=0.95\columnwidth]{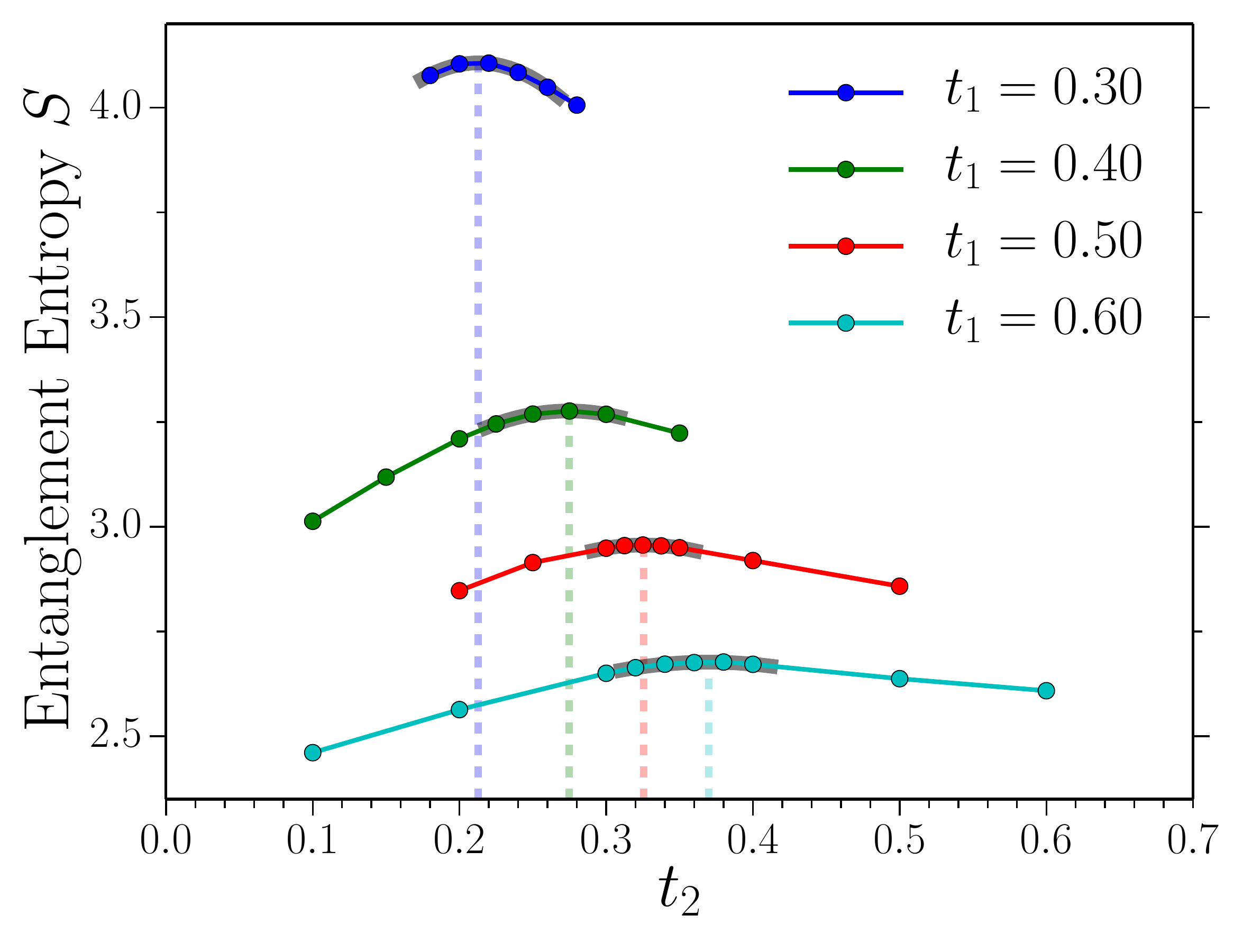}
\caption{Peaks in $N_y=8$ cylinder entanglement entropies as a function of $t_2$ for various fixed $t_1$ ($t_3=1$ throughout).
To estimate the peak locations, which in the 2D limit coincide with the transition out of the Fibonacci phase, data points near the peaks were fitted to a quadratic form.
Fits are shown as thick gray curves while the resulting peak locations
are indicated by vertical dashed lines. The peak locations give the four $N_y=8$ data points in Fig.~\ref{fig:phase_transition}.
The $t_1=0.4, 0.5, 0.6$ data were computed using DMRG with a fixed number of states $m$. The $t_1=0.3$ curve was found by extrapolating
fixed $t_2$ entropies as a power law in~$m$.
}
\label{fig:Ny8entropy}
\end{figure}

The previous subsection reported substantial evidence that the model in Eq.~\eqref{eqn:model} realizes the Fibonacci phase along the line $t_1 = t_2 \equiv t_\perp$, for a wide range of $t_\perp/t_3$.
In light of this finding it is interesting to now explore the extent of the Fibonacci phase for general $t_1, t_2, t_3 \geq 0$.
To address this question we fix the ratio $t_1/t_3$ and then vary $t_2$ from 0 to $t_1$, thus scanning a ray in parameter space.
Along this line we compute the bipartite entanglement entropy of infinite cylinders, observing clear peaks in the entropy as a function of $t_2$ (Fig.~\ref{fig:Ny8entropy}) that indicate a transition out of the Fibonacci phase (smoothed into a crossover due to finite-size effects; we provide supporting evidence for this interpretation below).
These peaks are quite broad for systems with $t_1/t_3 \approx 1$ but sharpen considerably as $t_1/t_3$ approaches zero, presumably because for small $t_{1,2}/t_3$
the $y$-direction correlation length remains below the cylinder circumference except for points very close to the phase boundary.

\begin{figure}[t]
\includegraphics[width=0.95\columnwidth]{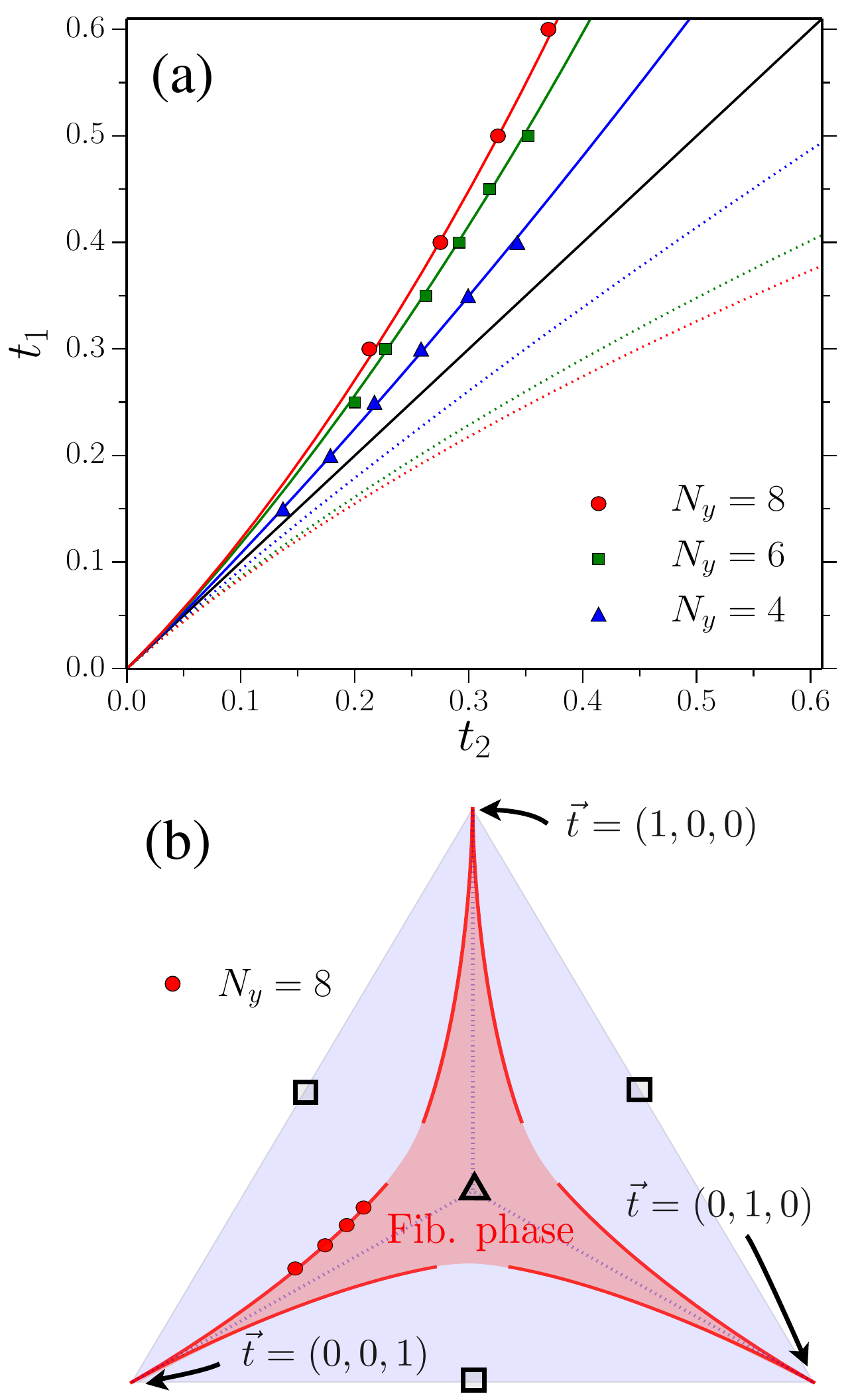}
\caption{(a) Estimated extent of the Fibonacci phase near the weakly-coupled-chain limit. Data points correspond to the locations of peaks in the entanglement entropy observed when
 tuning $0<t_2<t_1$ for fixed $t_1$ in $N_y=4,6,8$ cylinders ($t_3=1$ throughout). The solid curves are fits of the data to the function
 $(t_2-t_1)=C (t_2+t_1)^{8/5}$ with $C$ an $N_y$-dependent fitting parameter.
 Dashed curves show the fits reflected across the line $t_1=t_2$, since in the thermodynamic limit the Hamiltonian is symmetric about this line.
 (b) $N_y=8$ fit reflected under all $t_1$--$t_2$--$t_3$ permutations, estimating the full extent of the Fibonacci phase.  The black squares and triangle mark the
 location of the isotropic square- and triangular-lattice models.
}
\label{fig:phase_transition}
\end{figure}

To estimate the locations of the entropy peaks, data points near each peak were fit to a quadratic as shown in Fig.~\ref{fig:Ny8entropy} for the case of $N_y=8$ cylinders.
Figure~\ref{fig:phase_transition}(a) shows the locations of the entropy peaks thus obtained for $N_y=4,6,8$ cylinders.
For reference the black $t_1 = t_2$ curve represents the critical line for the two-chain limit discussed in Sec.~\ref{sec:two_chain}; we now see that, as expected, this line broadens into an extended phase on larger cylinders.
As a useful consistency check we note that for a 2D system composed of weakly coupled chains one can analytically constrain the shape of the phase boundary, which should precisely coincide with these entropy peaks extrapolated to the $N_y \rightarrow \infty$ limit.
In particular, scaling arguments reviewed in Appendix~\ref{appendix:CFT} predict that close to decoupled chains the critical couplings $t_{1c}$ and $t_{2c}$ satisfy
\begin{align}
t_{2c}-t_{1c} = C\,(t_{2c}+t_{1c})^{8/5},
\end{align}
where $C$ denotes a non-universal constant and we have employed units where $t_3=1$.
Fits of the data to this form appear in Fig.~\ref{fig:phase_transition}(a) (solid lines).
Despite having only one fitting parameter---and using data for moderately coupled chains---the fits are remarkably good even for the smaller cylinders.
The agreement motivated us to also fit the data to an arbitrary power law;
we find that for each $N_y$ the fitted exponent agrees with the predicted value of $8/5$ to within $10\%$.

\begin{figure*}[t]
\includegraphics[width=0.94\textwidth]{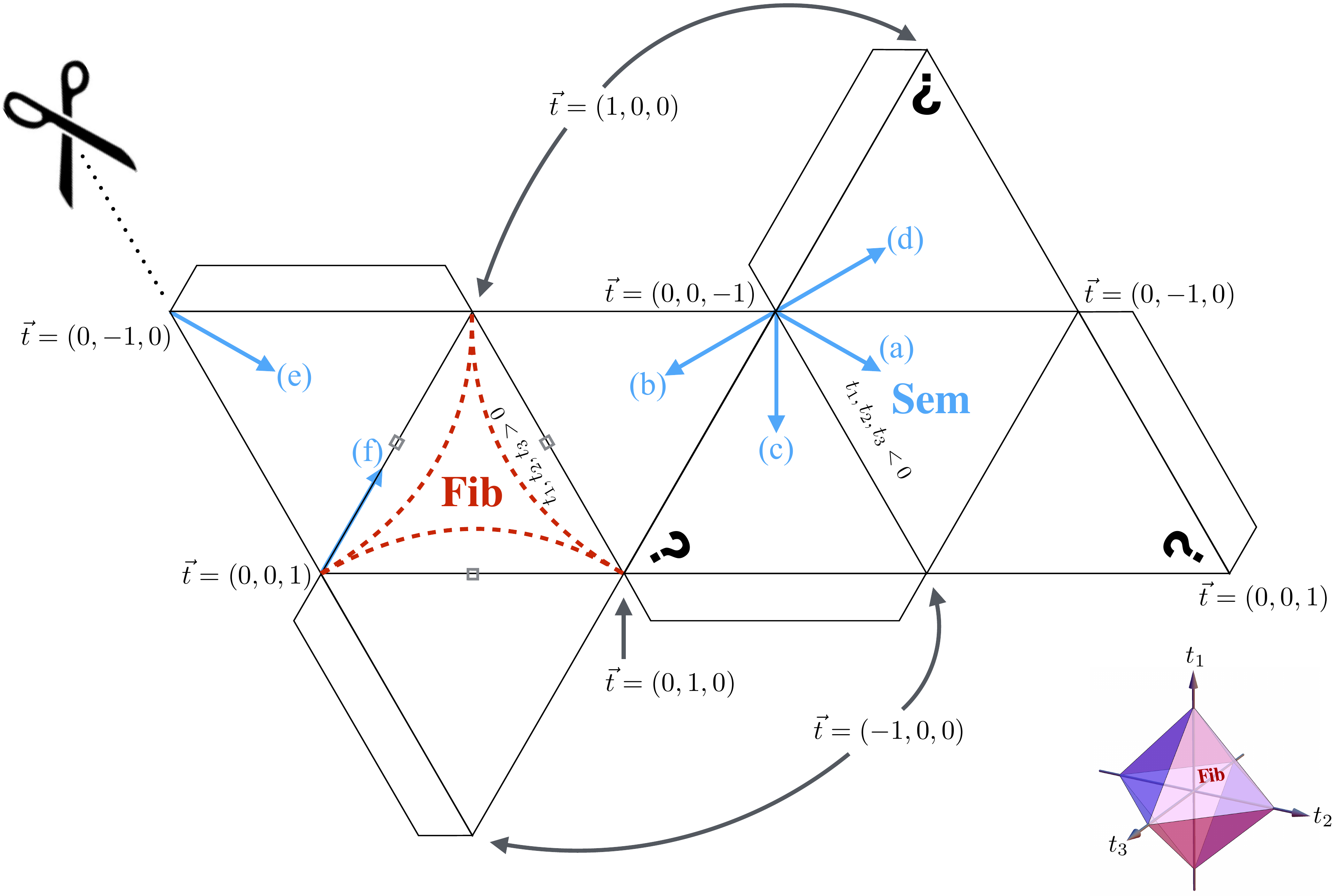}
\caption{
	Global phase diagram (some assembly required) of the Hamiltonian in Eq.~\eqref{eqn:model} with couplings constrained to \mbox{$|t_1|+|t_2|+|t_3| = 1$}.
	Note that Fig.~\ref{fig:phase_transition}(b) shows the face with $t_1, t_2, t_3 \geq 0$ containing the Fibonacci phase.
	The six corners of the octahedron correspond to various exactly solvable decoupled-chain limits, and directions (a)--(e) mark
	weakly-coupled chain limits where DMRG  shows strong evidence of the semion phase. Direction (f) indicates  anisotropic-square-lattice systems
	studied in Fig.~\ref{fig:sq_endiff} and small squares mark locations of the
	isotropic square lattice model in the 2D limit. \\
	Bottom right corner: Phase diagram after assembly; refer to \href{http://www.youtube.com/watch?v=kUVVQwtpX4o}{{\color{blue}\underline{YouTube}}} for instructions.
	}
\label{fig:phase_octahedron}
\end{figure*}

In the 2D limit, the phase diagram for Eq.~\eqref{eqn:model} must be symmetric under permutations of $t_1$, $t_2$, and $t_3$.
We can exploit this symmetry to roughly estimate the full extent of the Fibonacci phase for all $t_1, t_2, t_3 > 0$ based on the DMRG data discussed above.
As a first step, the dashed curves in Fig.~\ref{fig:phase_transition}(a) reflect our DMRG data about the $t_1 = t_2$ line.
We note, however, that finite-$N_y$ systems break the permutation symmetry; this naive reflection thus represents an approximation whose validity increases with $N_y$.
Ideally we could have directly measured entropy peaks near the expected phase transition curve on the $t_2 > t_1$ side also. But we found
 finite cylinders within this region to be much more highly entangled, preventing an accurate DMRG study.
We then further permute our $N_y = 8$ data to arrive at the phase diagram shown in Fig.~\ref{fig:phase_transition}(b), parametrized in terms of a vector
\begin{equation}
  \vec t = (t_1,t_2,t_3)
\end{equation}
with $t_1 + t_2 + t_3 = 1$ and all couplings non-negative.
Notably, the isotropic triangular lattice point sits deep within the center of the phase---consistent with our results from the previous subsection.

\begin{figure*}[tb]
	\begin{tabular}{ll}
	\quad\subfigure[]{\label{fig:Z2_es_parabola}}&
	\quad\subfigure[]{\label{fig:Z2_es_data}}
	\\[-1mm]
	\includegraphics[width=\columnwidth]{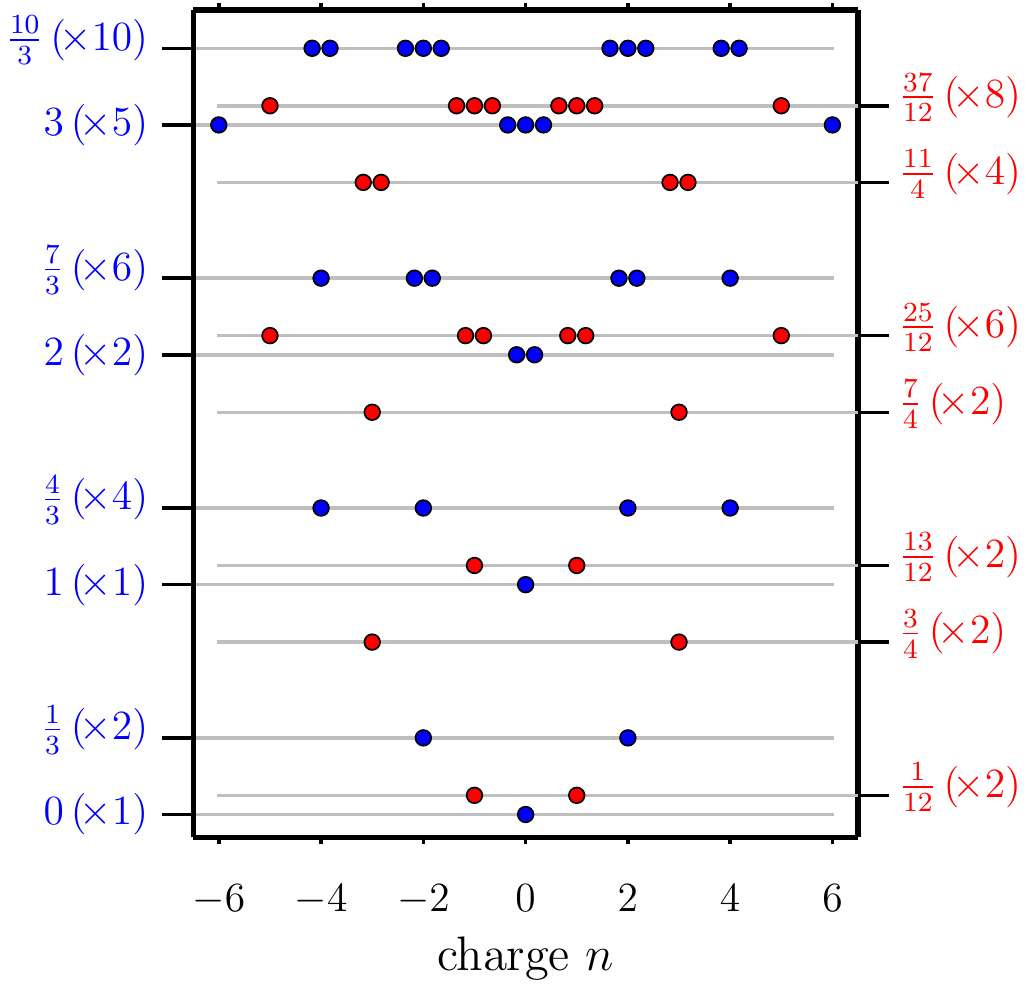}&
	\raisebox{43mm}{\begin{tabular}{c}\includegraphics[width=0.995\columnwidth]{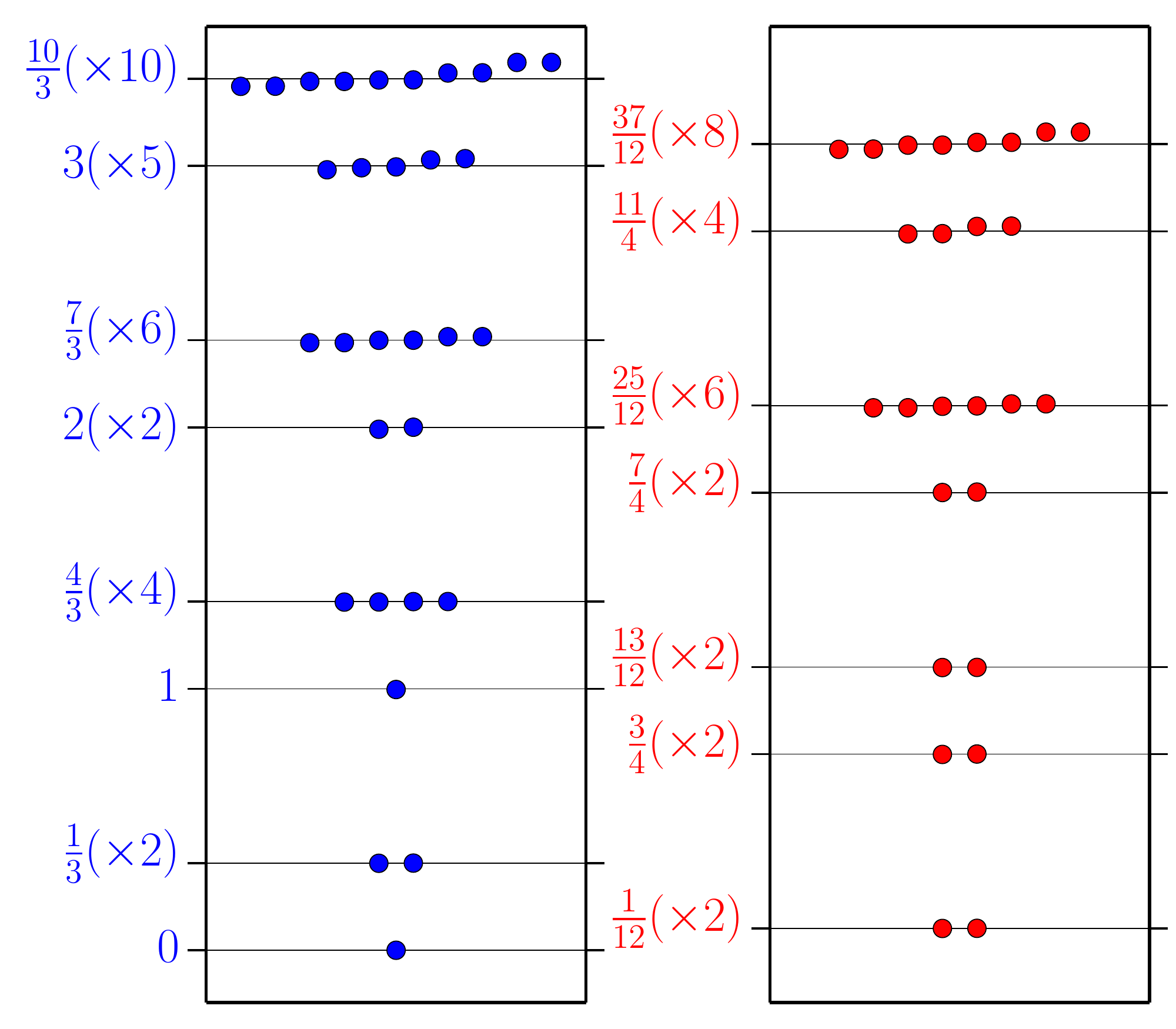} \\ \hspace{12mm}\mbox{\large $\ket{1_s}$}\hspace{37mm}\mbox{\large $\ket{s}$} \end{tabular}}
	\end{tabular}
	\caption{%
	(a) States of the $\mathrm{U(1)_6}$ chiral CFT, organized by charge $n$.  The highest weight states have energies $\frac{n^2}{12}$ and thus form a parabola.
		Their descendents have energies $\frac{n^2}{12}+m$ for integers $m=0,1,2,\ldots$ with multiplicities given by the partitions of the integers
		$1,1,2,3,\ldots$ \ .
		Combining all the states for even $n$ yields the edge spectra for the $\ket{1_s}$ ground state, shown on the left axis.
		The states for odd $n$ constitute the edge spectra for the $\ket{s}$ ground state, shown on the right axis.
	(b) Numerical entanglement spectra of the quasi-ground states
		$\ket{1_s}$ (blue, left column) and $\ket{s}$ (red, right column)
		at \mbox{$\vec{t}=(-0.1,-0.1,-1)$} for the $N_y=8$ cylinder after shifting and rescaling the spectra
        to match the lowest two levels of the $\mathrm{U(1)_6}$ CFT prediction.
        We used the same rescaling for both spectra but different shifts.  Numbers in parentheses denote the predicted degeneracies.
        The excellent agreement with analytics strongly supports the onset of the semion phase.}
	\label{fig:Z2_es}
\end{figure*}

\subsection{Global Phase Diagram of the \texorpdfstring{$t_1$--$t_2$--$t_3$}{t1-t2-t3} Model \label{sec:2d_phase_diagram}}

Our numerical results presented above indicate that the isotropic square lattice
(points in Fig.~\ref{fig:phase_transition}(b) marked by open squares) realizes a state distinct from the Fibonacci phase.
Pinpointing the precise nature of this state is, however, rather nontrivial.
Approaching the square lattice from the weakly-coupled-chain limit
\mbox{$0 < t_1/t_3 \ll 1$} with $t_2=0$ does not provide immediate insight, as the continuum limit of the $t_1$ coupling yields
competing relevant interactions whose effect is difficult to understand analytically.
Similarly, applying DMRG to small, anisotropic square-lattice cylinders
does not produce any clear evidence of spontaneous symmetry breaking or signatures of known topological phases.
States appearing when the model contains negative couplings also remain mysterious at this point, since throughout this section we have so far focused exclusively on regimes with $t_{1,2,3} \geq 0$.
If we allow for arbitrary signs of $t_{1,2,3}$, Fig.~\ref{fig:phase_transition}(b) actually contains only one out of \emph{eight} faces in the full phase diagram for the Hamiltonian in Eq.~\eqref{eqn:model}!
(For instance, a second face arises if we take $t_1 < 0$ but $t_{2,3}>0$ and so on; see Fig.~\ref{fig:phase_octahedron} for an `unfolded' sketch of the eight faces.)
In what follows we aim to fill in these gaps and deduce the ground states appearing in the broader parameter space.

Fortunately, we have not yet exhausted all of the analytically soluble windows: with negative couplings a new tractable weakly-coupled-chain limit emerges that lends a great deal of insight into the problem.
Consider the case $t_1 = t_2 = 0$ with negative intrachain coupling $t_3 < 0$.
Under a Fradkin-Kadanoff transformation\cite{FradkinKadanoff} each decoupled chain here maps to the self-dual point of the \emph{antiferromagnetic} three-state Potts model,
	which is described by a non-chiral $\mathrm{U(1)}_6$ CFT with $K$-matrix $K = \left(\begin{smallmatrix}6&\\&-6\end{smallmatrix}\right)$ and central charge $c = 1$.\cite{CardyJacobsenSokal:AF3Potts}
We respectively denote the corresponding left- and right-moving chiral boson fields for chain $y$ by $\phi_{L}(y)$ and $\phi_{R}(y)$, both $2\pi$-periodic in our normalization conventions.

Just as with positive $t_3$, we can once again bootstrap off of these decoupled critical chains to deduce the system's response to small but finite $t_{1,2}$.
To do so we must expand the interchain couplings in Eq.~\eqref{eqn:model} in terms of low-energy fields captured by the CFT.
Reference~\onlinecite{Delfino:AFPotts:2001} already provided the desired mapping between lattice operators and continuum fields for the self-dual antiferromagnetic Potts model;
the result reads
\begin{align}
  \alpha_{R/L} \sim e^{2i\phi_{R/L}} + \cdots
  \label{AF_alpha_expansion}
\end{align}
where the ellipsis denotes subdominant contributions.
Using this mapping, 
the interchain Hamiltonian $H_\perp$ takes the form
\begin{align}\begin{split}
	H_\perp &\sim \sum_y \int_x \Big[ (t_1+t_2) e^{2i\phi_R(y)} e^{-2i\phi_L(y+1)} + H.c.
	\\	&\qquad	+ (t_1-t_2) X(\phi_{L,R}(y),\phi_{L,R}(y+1)) \Big] .
	\label{eq:AF_perturbations}
\end{split}\end{align}
The first term possesses scaling dimension $\Delta=\frac23$ at the decoupled-chain fixed point.  In the second line $X$ represents (numerous) subleading terms with dimension $\Delta=\frac53$; these can be enumerated explicitly but we refrain from doing so here.

Except with fine-tuning, the 2D system's properties are generically dominated by the more relevant $(t_1+t_2)$ term above.
Its effect is relatively easy to understand for systems composed of a large but finite number of chains.
When $(t_1+t_2)\neq0$, the first line in $H_\perp$ directly couples the right-mover of chain $y$ with the left-mover of $y+1$.  This hybridization gaps out the entire bulk but leaves `unpaired' \emph{chiral} left- and right-moving $\mathrm{U(1)_6}$ fields at the first and last chains, respectively.
The boundary CFT thus admits chiral central charge $c=-1$ and contains primary fields $e^{in\phi}$ with scaling dimensions $\Delta = \frac{n^2}{12}$ and spins $e^{-i\pi n^2/6}$ ($\phi$ is the gapless $2\pi$-periodic chiral boson field at the edge and $n$ is an integer `charge').

Note that these conclusions hold independent of the sign of $(t_1+t_2)$.

Crucially, this boundary CFT does not describe the edge with the vacuum since the lattice parafermion operators comprising our system reside in the interior of a fractionalized quantum Hall fluid.
Some care is therefore necessary when using the bulk-boundary correspondence to deduce properties of the gapped phase we have entered.
All fields in the boundary CFT must correspond \emph{either} to bulk quasiparticles originating from the coupled parafermions of interest \emph{or} quasiparticles native to the surrounding quantum Hall fluid.
To determine the former we must therefore `mod out' by the latter.
For concreteness, consider the first chain where $\phi\equiv\phi_L(1)$ describes the gapless degress of freedom.
Parafermion trilinears $\alpha_L\alpha_L\alpha_L$ are bosonic and thus $e^{6i\phi}$ is local given our operator identification in Eq.~\eqref{AF_alpha_expansion}.

While single parafermion operators $\alpha_L$ are clearly not bosonic, their nontrivial statistics originates from the fractionalized bulk quasiparticle of the parent quantum Hall fluid; in this sense $\alpha_L \sim e^{2i\phi}$ should also be considered `local'.
The fields $e^{2in\phi}$ for $n\in\mathbb{Z}$ therefore correspond to the quasiparticle content of the parent phase.
Such fields are all related by some `local' operator and belong to the same superselection sector;
likewise, the fields with odd $n$ form another superselection sector.
One can succinctly organize the field content---modulo the trivial boson $e^{6i\phi}$---into the sets
  $\{1,e^{2i\phi},e^{4i\phi}\}$ and $\{e^{i\phi},e^{3i\phi},e^{5i\phi}\}$.
Modding out further by the `local' field $e^{2i\phi}$ leaves only $1$ and $e^{3i\phi}$, corresponding to the quasiparticles arising from our coupled parafermions.
We conclude that for weakly coupled chains with $t_3 < 0$, the $(t_1+t_2)$ term in Eq.~\eqref{eq:AF_perturbations} drives the system into a topologically ordered
phase with a two-fold ground state degeneracy on a torus or infinite cylinder.
Fusing the boundary field $e^{3i\phi}$ with itself yields $e^{3i\phi} \times e^{3i\phi} \sim e^{6i\phi}$, a trivial boson, hence the nontrivial anyon has quantum dimension $d=1$.
Moreover, since the spin of the boundary field $e^{3i\phi}$

is $e^{-9i\pi/6} = i$, this nontrivial bulk anyon is a semion.%
	\footnote{We could alternatively have used (say) $e^{i\phi}$ instead of $e^{3i\phi}$ for the boundary field; both yield the same spin modulo the spin for the parent quantum Hall quasiparticles, however, so our conclusions are insensitive to this choice.}
We will therefore refer to this Abelian state as the ``semion phase''.

We confirm the presence of the semion phase by first applying infinite-cylinder DMRG to the Hamiltonian with \mbox{$\vec{t} = (-0.1,-0.1,-1)$},
which lies along the direction labeled~(a) in Fig.~\ref{fig:phase_octahedron}.
Starting from random initial states our simulations indeed recover two quasi-ground states that we denote $\ket{1_s}$ and $\ket{s}$.
(The `$s$' subscript on the former is used to distinguish from the state $|1\rangle$ defined for the Fibonacci phase.)
Their entanglement spectra, shown in Fig.~\ref{fig:Z2_es_data}, agree well with the $\mathrm{U(1)_6}$ CFT counting shown in Fig.~\ref{fig:Z2_es_parabola}.
We also find very similar results (not shown) for weakly-coupled-chain systems with \mbox{$t_1=t_2 > 0$}
along direction~(b) of Fig.~\ref{fig:phase_octahedron}, and for systems along direction~(e).
Note that in the 2D limit directions~(b) and (e) are related under
permutation of $t_2$ and $t_3$.

The less-relevant $(t_1-t_2)$ term in Eq.~\eqref{eq:AF_perturbations} plays an important role only when $(t_1 + t_2)$ is tuned near zero.
As mentioned above, this part of the Hamiltonian contains a sum of competing terms, each with scaling dimension $\frac53$,
making it difficult to determine analytically which 2D phase it favors.
Nevertheless, by turning again to numerics we find that along the line $t_1=-t_2$  close to the \mbox{$\vec{t}=(0,0,-1)$} point [directions (c) and (d) in Fig.~\ref{fig:phase_octahedron}]
the system still exhibits two quasi-ground states with entanglement spectra consistent with the semion phase.
This observation---along with the results for directions (a), (b), and (e)---leads us to believe that the semion phase occupies a large swath of parameter space around the symmetry-related points
$\vec{t} = (0,0,-1)$, $\vec{t} = (0,-1,0)$, and $\vec{t} = (-1,0,0)$.

\begin{figure}
\includegraphics[width=\columnwidth]{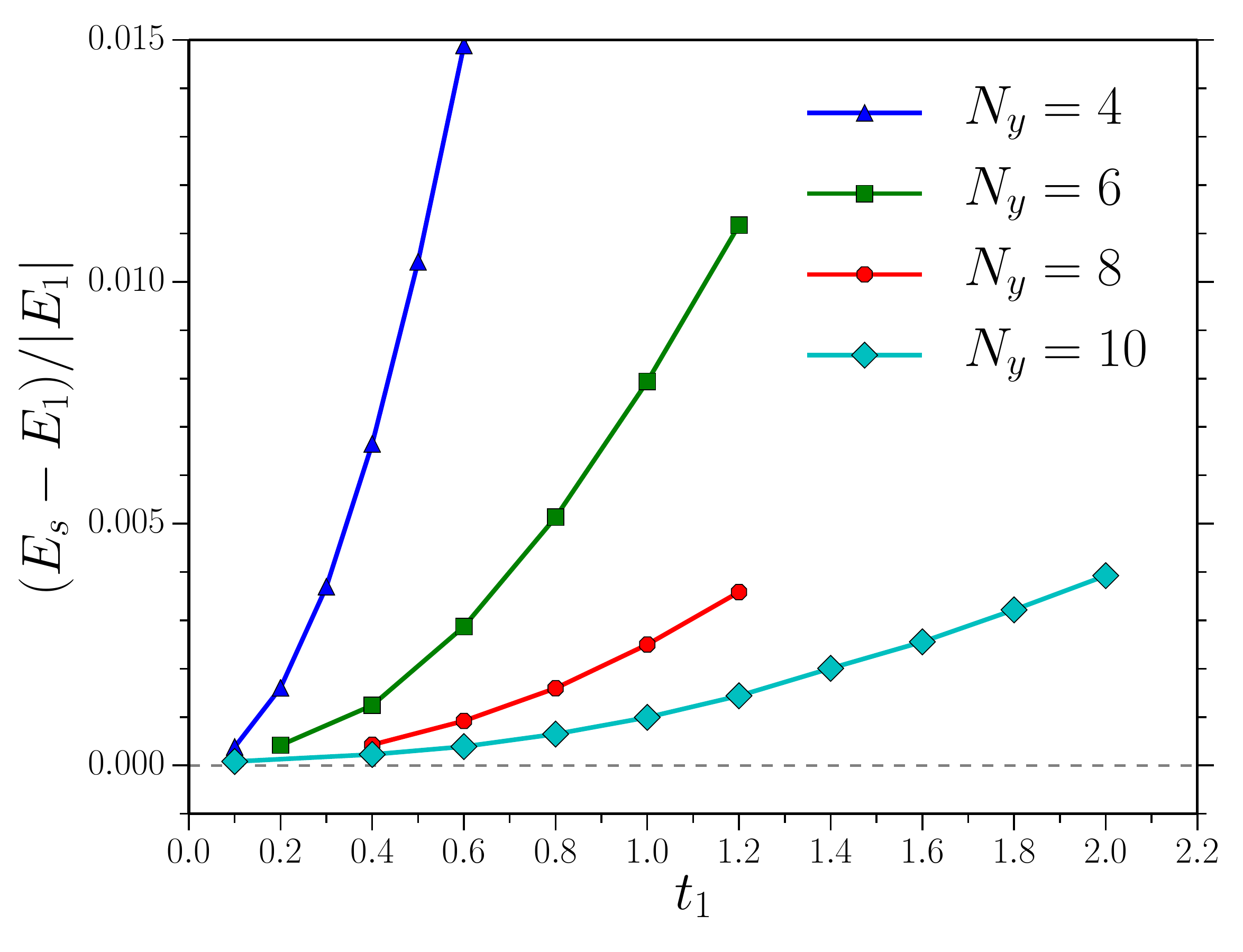}
\caption{Relative energy splitting between quasi-degenerate ground states of cylinders with $N_y=4$--$10$  as a function of $t_1$ (in units where $t_3=1$ with $t_2=0$ throughout).
	The horizontal axis corresponds to direction (f) in Fig.~\ref{fig:phase_octahedron}.}
\label{fig:sq_endiff}
\end{figure}

Let us now return to the question of which phase emerges in the isotropic
square lattice limit with positive couplings, that is, $\vec{t}=(1,0,1)$ and its permutations. Directly approaching the $\vec{t}=(1,0,1)$ point from the
decoupled-chain limit along direction (f) in Fig.~\ref{fig:phase_octahedron} is unfortunately challenging for DMRG.
Here the $y$-direction correlation length seems to grow very quickly as a function of $t_1/t_3$ relative to the slower
growth encountered for the anisotropic triangular case upon varying $t_\perp/t_3$. As indirect evidence of the growth of $y$ correlations,
Fig.~\ref{fig:sq_endiff} shows the relative energy splitting of the two anisotropic-square lattice quasi-ground states
as a function of $t_1$ (with $t_2=0$ and $t_3=1$). Compared to Fig.~\ref{fig:tperp_endiff} for the anisotropic triangular lattice,
 the anisotropic-square-lattice ground states separate more rapidly with $t_1$ and remain significantly split up to larger system sizes.
Nevertheless, we consistently observe two quasi-ground states on anisotropic square lattice cylinders up through
the isotropic point $t_1=t_3=1$ (however, we cannot rule out the possibility that more than two exist).
Combined with our strong evidence that the semion phase appears along directions (b) and (e) in Fig.~\ref{fig:phase_octahedron} and persists in a broad region around \mbox{$\vec{t}=(0,0,-1)$} and symmetry-related points, we find it reasonable to conjecture that the isotropic square-lattice system
 also lies within the semion phase.

Finally, we briefly comment on the symmetry-related regions labeled with question marks in Fig.~\ref{fig:phase_octahedron}.
These parameter regimes represent weakly coupled chains described by $\Z3$ parafermion CFT's hybridized via \emph{negative} interchain couplings.
Consider the special case where $t_3 = 1$ with small $t_1 = t_2 \equiv t_\perp < 0$.
Here the most relevant term generated by $t_\perp$ corresponds to the perturbation that drives the $c = 4/5$ parafermion CFT to the tricritical Ising theory with $c = 7/10$.\cite{Klassen:DSeriesCFTFlow:1993}
We believe that the resulting critical state is unstable within our model, but one would need to assess the behavior of other less relevant terms to draw any conclusions.
The situation should be contrasted to the case with $t_\perp > 0$ where the leading perturbation opens a gap directly and yields the Fibonacci phase.
Determining the fate of the system here is therefore expected to be especially delicate, so we leave an exploration of this subtle regime to future work.

\section{Conclusions}
\label{Conclusions}

Coupling arrays of extrinsic defects---e.g., those that bind parafermionic zero modes---provides a promising avenue towards realizing exotic phases of matter, particularly given possible realizations in quantum-Hall architectures.
In an early work, Burrello et al.\cite{Burrello:2013} constructed a model of coupled parafermions supporting an Abelian topologically ordered state that generalizes the toric code.
Reference~\onlinecite{Mong:2014} explored an alternative setup consisting of weakly coupled parafermion chains that were analytically shown to enter a `Fibonacci phase' with non-Abelian Fibonacci anyons that possess universal braid statistics.
This study established proof of principle that one could assemble hardware for a universal topological quantum computer using well-understood Abelian phases of matter.
We stress, however, that the quasi-1D limit invoked there served purely as a theoretical crutch.
Exploring the broader stability of the Fibonacci phase away from this soluble regime is therefore important for assessing the utility of this line of attack.
Here we employed extensive DMRG simulations, bolstered by complementary analytical results, to address precisely this issue in a parafermion model that contains decoupled chains and the isotropic square- and triangular-lattices as special cases.

Reassuringly, our numerics reveal a remarkably resilient Fibonacci phase---evidenced by the telltale ground-state degeneracy, entanglement entropy, and entanglement spectrum---extending well beyond the weakly-coupled-chain limit; recall Fig.~\ref{fig:phase_transition}(b).
Elsewhere in the phase diagram an Abelian topologically ordered state with semion excitations appears.
Although we have not exhaustively studied the model's parameter space, our simulations revealed only these two phases (which, curiously, comprise the only two topological orders with a single nontrivial quasiparticle type).
Note that we also performed simulations of square-lattice systems with first- and second-neighbor interchain couplings, though for brevity the results were not presented.
This alternative setup arises by augmenting Fig.~\ref{fig:triangular_lattice} with upward-sloping diagonal bonds.
We found a clear Fibonacci phase here as well, which nicely emphasizes the robustness of the physics against changes in microscopic details.

It is worth re-emphasizing the exceedingly useful complementarity of coupled-chain analytics and our DMRG simulations.
The quasi-1D limit predicts sharp fingerprints of the Fibonacci phase that, because of the ultra-short transverse correlation lengths characteristic of this regime, could be verified numerically even in systems composed of as few as four chains.
This feature allowed us to controllably track the extent of the Fibonacci phase in regimes where we lack analytical control---which of course was the primary goal of our simulations.
We expect that analogous quasi-1D deformations should be useful for numerics in a variety of other contexts as well.
One other aspect of the model is also worth stressing:  it consists entirely of quadratic near-neighbor parafermion couplings yet captures highly exotic topological phases of matter; this is possible because even with only bilinear terms the Hamiltonian is strongly interacting.\cite{Fendley:2012}
The model's simplicity is certainly a boon for the prospect of realizing Fibonacci anyons in the underlying quantum-Hall setups.

Our results also yield indirect implications for systems built out of local bosonic degrees of freedom.
In an interesting recent study, Barkeshli et al.\cite{BarkeshliFib} introduced generalized Kitaev honeycomb models for which the constituent spin operators could be decomposed in terms of `slave parafermions'.
Within that representation the authors identified a quasi-1D limit that also realizes a phase with Fibonacci anyons, among other quasiparticles.
Reference~\onlinecite{Vaezi:2014b} conjectured the appearance of Fibonacci anyons in related local bosonic models (though Abelian topological orders were found in a mean-field treatment).
By applying the `inverse' logic from these works one can back out local bosonic setups corresponding to the parafermion system that we analyzed.
Doing so immediately allows one to construct bosonic lattice models supporting a phase with Fibonacci anyons---though the Hamiltonians involve somewhat elaborate multi-boson interactions.
It would be interesting in future numerical work to explore the extent of that state when the model is deformed towards a simpler limit with only quadratic couplings.

One particularly intriguiging wide-open question is whether the Fibonacci phase can appear in a reasonable \emph{spatially uniform} Hamiltonian describing Abelian quantum Hall/superconductor heterostructures or bilayers (i.e., with the periodic modulations trapping the parafermion modes `smeared out').
If found, such setups might offer relatively simple experimental routes to Fibonacci anyons.
The fact that we identified an isotropic triangular lattice model supporting the Fibonacci phase is encouraging and may offer some clues as to how one can construct spatially uniform analogues.
We also note that Barkeshli and Vaezi\cite{Vaezi:2014} proposed remarkably simple $\nu = \frac13+\frac13$ bilayers as uniform Fibonacci anyon hosts based on an analysis of the thin-torus limit.
A concerted numerical push in this direction seems a very worthwhile complement to the lattice simulations performed here.

\section*{Acknowledgments}
E.~M.~S.~acknowledges helpful discussions with Lukasz Cincio and Juan Carrasquilla.
Simulations were performed on the Perimeter Institute HPC.
We are grateful for generous support from the Sherman Fairchild Foundation (R.~M.), the National Science Foundation through grant DMR-1341822 (J.\ A.); the Alfred P.\ Sloan Foundation (J.\ A.); the Caltech Institute for Quantum Information and Matter, an NSF Physics Frontiers Center with support of the Gordon and Betty Moore Foundation through Grant GBMF1250; and the Walter Burke Institute for Theoretical Physics at Caltech.

\appendix

\section{Mapping to Potts chain}
\label{appendix:Potts_mapping}

Here we very briefly review the mapping of a 1D $\Z3$ parafermion chain to the three-state quantum Potts model.\cite{Fendley:2012,FradkinKadanoff}
Consider the Hamiltonian in Eq.~\eqref{eqn:single_trench}, repeated here for clarity,
\begin{align}
  H = - \sum_n & \Bigl[ \omega\, ( J\, \alpha^\dagger_{R,2n+1} \alpha_{R,2n} + h\, \alpha^\dagger_{R,2n} \alpha_{R,2n-1} ) \nonumber \\
  & \mbox{} +  \text{H.c.} \Bigr] \label{eqn:single_trench2}  \:,
\end{align}
where again $\omega = e^{i(2\pi/3)}$.
We have chosen to express (somewhat arbitrarily) the Hamiltonian using `right' parafermion operators that satisfy $\alpha_{R,j}^3=1$, $\alpha_{R,j}^\dagger=\alpha_{R,j}^2$, and
\begin{align}
  \alpha_{R,j} \alpha_{R,j^\prime} & = e^{i (2\pi/3)\, \text{sgn}(j^\prime-j) } \alpha_{R,j^\prime} \alpha_{R,j}.
\end{align}
One could equally well employ `left' parafermion operators whose commutator instead reads
\begin{align}
  \alpha_{L,j} \alpha_{L,j^\prime} & = e^{-i (2\pi/3)\, \text{sgn}(j^\prime-j) } \alpha_{L,j^\prime} \alpha_{L,j}.
\end{align}

In either representation, the mapping to the Potts model proceeds upon nonlocally decomposing the parafermions via
\begin{eqnarray}
  \alpha_{R,2n-1}&=&\sigma_n\prod_{m<n}\tau_m,~~~~\alpha_{R,2n}=\omega\sigma_n\prod_{m\leq n}\tau_m
  \label{R}  \label{eqn:Rmapping} \\
  \alpha_{L,2n-1}&=&\sigma_n\prod_{m<n}\tau_m^\dagger,~~~~\alpha_{L,2n}=\omega^{2}\sigma_n\prod_{m\leq n}\tau_m^\dagger
  \label{L}. \ \label{eqn:Lmapping}
\end{eqnarray}
On the right sides we have introduced local bosonic Potts operators $\sigma_n, \tau_n$ that satisfy
\begin{eqnarray}
  \sigma_n^3 = \tau_n^3 = 1, ~~~~\sigma_n^\dagger = \sigma_n^2, ~~~~\tau_n^\dagger = \tau_n^2
\end{eqnarray}
together with the commutator
\begin{equation}
  \sigma_n\tau_n = \omega \tau_n\sigma_n.
\end{equation}
(In contrast to the parafermion operators, $\sigma_m$ and $\tau_n$ commute for $m \neq n$.)
Notice that the right and left representations differ primarily in the form of their non-local $\tau_m$ strings.
One can readily verify that the decompositions in Eq.~\eqref{R} and \eqref{L} preserve the parafermion operator algebra.
Inserting these expressions into Eq.~\eqref{eqn:single_trench2} yields the desired quantum Potts Hamiltonian,
\begin{equation}
H=-\sum_n\left[J\left(\sigma_{n+1}^\dagger\sigma_n + {\rm H.c.}\right) + h(\tau_n+\tau_n^\dagger)\right].
\end{equation}

\section{DMRG path \label{appendix:dmrg_path}}

When working with a model of coupled parafermions such as Eq.~\eqref{eqn:model}, it is necessary to specify
the parafermion operator commutation relations by ordering the sites of the system.
Such an ordering also specifies how the parafermion operators map into clock-model operators
under a Fradkin-Kadanoff transformation.

As an example, for a 2D array of parafermion operators a common choice of ordering is for operators on chain $y$ to precede those on chain \mbox{$y^\prime > y$},
while on the same chain operators remain ordered from left to right as in the 1D case.
For parafermion operators on the same chain~$y$ this implies
\begin{align}
\alpha_{R,j}(y) \alpha_{R,j^\prime}(y) & = e^{i (2\pi/ 3)\: \text{sgn}(j^\prime-j)} \alpha_{R,j^\prime}(y) \alpha_{R,j}(y) \nonumber \\
\alpha_{L,j}(y) \alpha_{L,j^\prime}(y) & = e^{-i (2\pi/ 3)\: \text{sgn}(j^\prime-j)} \alpha_{L,j^\prime}(y) \alpha_{L,j}(y) \ ,
\end{align}
while for operators on different chains,
\begin{align}
\alpha_{R,j}(y) \alpha_{R,j^\prime}(y^\prime) & = e^{i (2\pi/ 3)\: \text{sgn}(y^\prime-y)} \alpha_{R,j^\prime}(y^\prime) \alpha_{R,j}(y) \nonumber \\
\alpha_{L,j}(y) \alpha_{L,j^\prime}(y^\prime) & = e^{-i (2\pi/ 3)\: \text{sgn}(y^\prime-y)} \alpha_{L,j^\prime}(y^\prime) \alpha_{L,j}(y)
\end{align}
regardless of the order of $j$ and $j^\prime$.
When simulating a 2D system using DMRG, which requires choosing a 1D path for the matrix-product-state wavefunction,
it is convenient to choose the DMRG path to follow the parafermion ordering. However, for DMRG to follow the above
ordering would lead to a very steep growth in computational cost for systems with a significant number of sites along the $x$~direction.
Fortunately, however, other physically equivalent parafermion orderings are possible.

\begin{figure}
\includegraphics[width=0.8\columnwidth]{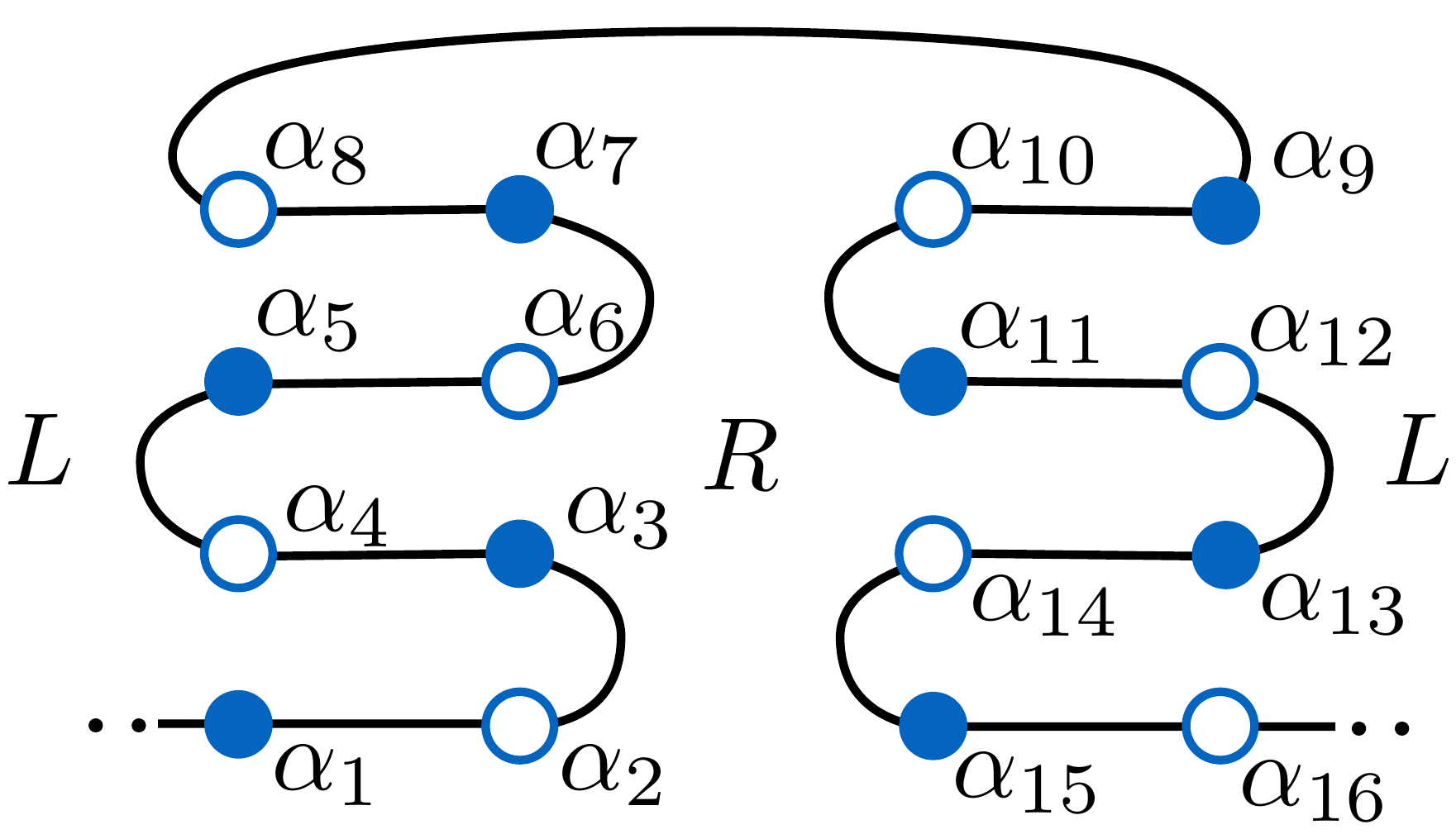}
\caption{Possible ordering of parafermion operators in an $N_y=4$ cylinder.  For the Hamiltonian in Eq.~\eqref{eqn:model}, this choice yields no interactions that cross
the parafermion ordering path.
Parafermion `hopping' processes that pass through regions labeled $R$ or $L$ are expressed in terms of the corresponding operator representation, e.g., $\alpha_{3R}^\dagger \alpha_{6R}$ or $\alpha_{5L}^\dagger \alpha_{8L}$.
}
\label{fig:alt_dmrg_ordering}
\end{figure}

The alternative ordering (and corresponding DMRG path) shown in Fig.~\ref{fig:alt_dmrg_ordering} is reasonably efficient for performing DMRG calculations.
It also makes implementing a Hamiltonian such as Eq.~\eqref{eqn:model} relatively simple since none of the interactions cross the parafermion path.
To map an interaction involving, say, sites 3 and 6 into clock-operator form, one uses the mapping Eq.~\eqref{eqn:Rmapping}, using the $R$
representation as this process `hops' parafermions \emph{under} the path (through the region labeled $R$ in Fig.~\ref{fig:alt_dmrg_ordering}).
The resulting correspondence is
\begin{align}
\omega \alpha^\dagger_{R,3} \alpha_{R,6} & = \sigma^\dagger_2 \tau_2 \tau_3 \sigma_3 \,.
\end{align}
The factor of $\omega$ is included above so that the Hamiltonian term involving this bilinear
$(\omega \alpha^\dagger_{R,3} \alpha_{R,6} + \text{H.c.})$ is symmetric under charge conjugation, defined as the mapping:
\begin{align}
\alpha_{R/L} & \leftrightarrow \alpha_{R/L}^\dagger
\label{eqn:cc} \, .
\end{align}

\begin{figure}
\includegraphics[width=0.6\columnwidth]{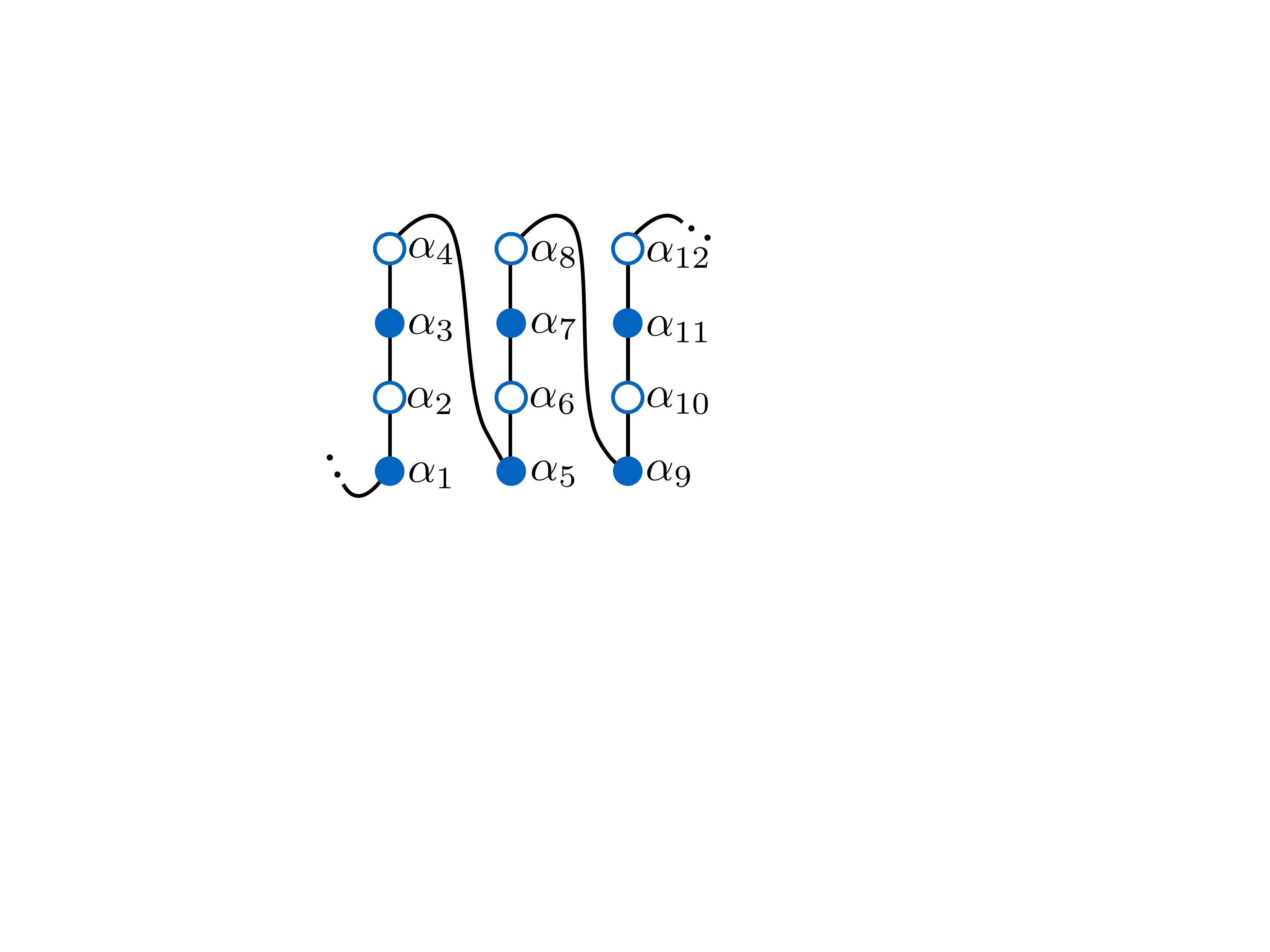}
\caption{Ordering of parafermion operators used in DMRG for the present work (shown for a cylinder with $N_y=4$).
Note that in terms of the underlying clock degrees of freedom seen by DMRG, this path requires a circumference of only two sites in the $y$ direction. }
\label{fig:dmrg_ordering}
\end{figure}

However, because of the way short-range interactions in 2D map to longer-ranged interactions in the 1D
path used by DMRG, the most efficient path choice for DMRG corresponds to ordering the sites by columns such that all operators in column $x$ are ordered before those in $x+1$.
Therefore in the present work we we have used the ordering illustrated in Fig.~\ref{fig:dmrg_ordering}.
Because such a path `pairs' parafermions into clock sites along the $y$-direction, it has the added benefit of reducing the number of clock sites
along the $y$-direction by a factor of two compared with a path that pairs parafermions into clock sites along the $x$-direction.

\begin{figure}
\includegraphics[width=0.6\columnwidth]{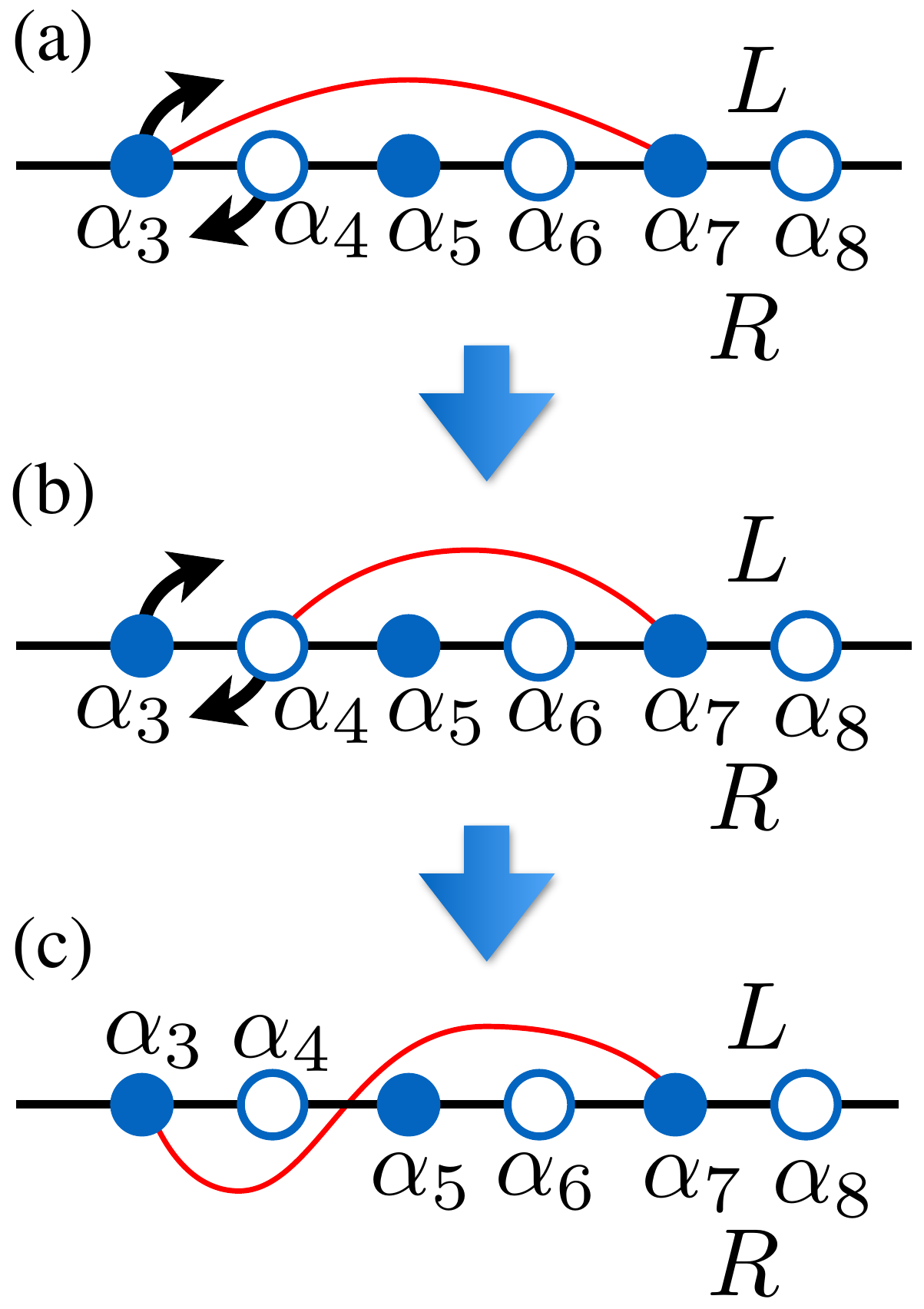}
\caption{`Braiding' parafermion sites 3 and 4 clockwise two times transforms the pairwise interaction between sites 3 and 7, defined originally in the
$L$ representation, into an interaction that crosses the parafermion ordering path.
This scheme allows us to simulate the model of Eq.~\eqref{eqn:model} using the numerically very convenient ordering shown in Fig.~\ref{fig:dmrg_ordering}.}
\label{fig:braid}
\end{figure}

Relative to the conventions specified in Fig.~\ref{fig:alt_dmrg_ordering}, implementing the model of Eq.~\eqref{eqn:model} using the ordering in Fig.~\ref{fig:dmrg_ordering} is subtler since here many Hamiltonian terms involve processes that cross the parafermion path (e.g., coupling between sites 3 and 7).
The clock-operator form of such processes can be derived from a simpler `reference' parafermion ordering in which the pairwise interaction involving the same two sites does not cross the
ordering path.
One way to carry out this derivation is to apply a unitary transformation that performs a clockwise `braid' of the parafermion sites \mbox{$(j,j+1)$} as follows
\begin{align}\begin{split}
  \alpha_{R,j} &\rightarrow \alpha_{R,j+1} \,, \\
  \alpha_{R,j+1} &\rightarrow \omega^2 \alpha_{R,j+1}^\dagger \alpha_{R,j}^\dagger   \label{eqn:Rbraidrules}
\end{split}\end{align}
for parafermionic operators in the $R$ representation; for those in the $L$ representation 
\begin{align}\begin{split}
  \alpha_{L,j} &\rightarrow \alpha_{L,j+1} \,, \\
  \alpha_{L,j+1} &\rightarrow \omega \alpha_{L,j+1}^\dagger \alpha_{L,j}^\dagger  \,. \label{eqn:Lbraidrules}
\end{split}\end{align}
The above mapping corresponds to the braid transformation derived for parafermionic zero modes in Ref.~\onlinecite{Clarke:2013a}.

As an illustrative example, consider the interaction `hopping' a parafermion from site 7 to site 3 in Fig.~\ref{fig:dmrg_ordering}, which crosses the ordering path.
We can construct such an interaction beginning from the simpler process depicted in Fig.~\ref{fig:braid}(a) that hops a parafermion from site 7 to site 3 while staying
above the path, thus involving parafermion operators in the $L$ representation:
\begin{align}
\omega \alpha^\dagger_{L,3} \alpha_{L,7} = \omega \sigma^\dagger_2 \tau^\dagger_2 \tau^\dagger_3 \sigma_4  \label{eqn:pfn_interaction} \ .
\end{align}
(We display this operator in both parafermionic and clock-operator form for clarity.)
Braiding sites 3 and 4 clockwise maps \mbox{$\alpha_{L3} \rightarrow \alpha_{L,4}$}, transforming this interaction into the one depicted
in Fig.~\ref{fig:braid}(b) as follows:
\begin{align}\begin{split}
\omega \alpha^\dagger_{L,3} \alpha_{L,7} & \rightarrow  \omega \alpha^\dagger_{L,4} \alpha_{L,7}  \\
& = \sigma^\dagger_2 \tau^\dagger_3 \sigma_4 \ .
\end{split}\end{align}
Finally, braiding sites 3 and 4 clockwise once more maps $\alpha_{L,4} \rightarrow \omega \alpha_{L,4}^\dagger \alpha_{L,3}^\dagger$,
yielding
\begin{align}\begin{split}
\omega \alpha_{L,4}^\dagger \alpha_{L,7} & \rightarrow \omega (\omega  \alpha_{L,4}^\dagger \alpha_{L,3}^\dagger)^\dagger \alpha_{L,7} \\
& = \alpha_{L,3} \alpha_{L,4} \alpha_{L,7}  \\
& = \omega^2 \sigma^\dagger_2 \tau_2 \tau^\dagger_3 \sigma_4 \label{eqn:dmrg_clock_op}
\end{split}\end{align}
The key result here is the last line above, which gives the clock-operator form of the interaction Eq.~(\ref{eqn:pfn_interaction}), but
consistent with the ordering convention of Fig.~\ref{fig:dmrg_ordering}.

Because the Fig.~\ref{fig:dmrg_ordering} path is a less natural, but numerically very convenient, choice for the Hamiltonian Eq.~\eqref{eqn:model} in that interactions cross the parafermion path, we confirmed that our mapping to clock operators passed a number of consistency checks.
One of these is that parafermion bilinears which do not intersect should commute. Another is that the resulting clock-operator forms of each
interaction term should respect charge-conjugation symmetry.
We also checked that implementing the Hamiltonian Eq.~\eqref{eqn:model} via both the Fig.~\ref{fig:alt_dmrg_ordering} and Fig.~\ref{fig:dmrg_ordering} paths
gives the same ground-state energy in DMRG for a wide range of system sizes and Hamiltonian parameters.

\section{Self-duality in the Two-leg Ladder \label{appendix:ladder_duality}}

\begin{figure}
\includegraphics[width=0.7\columnwidth]{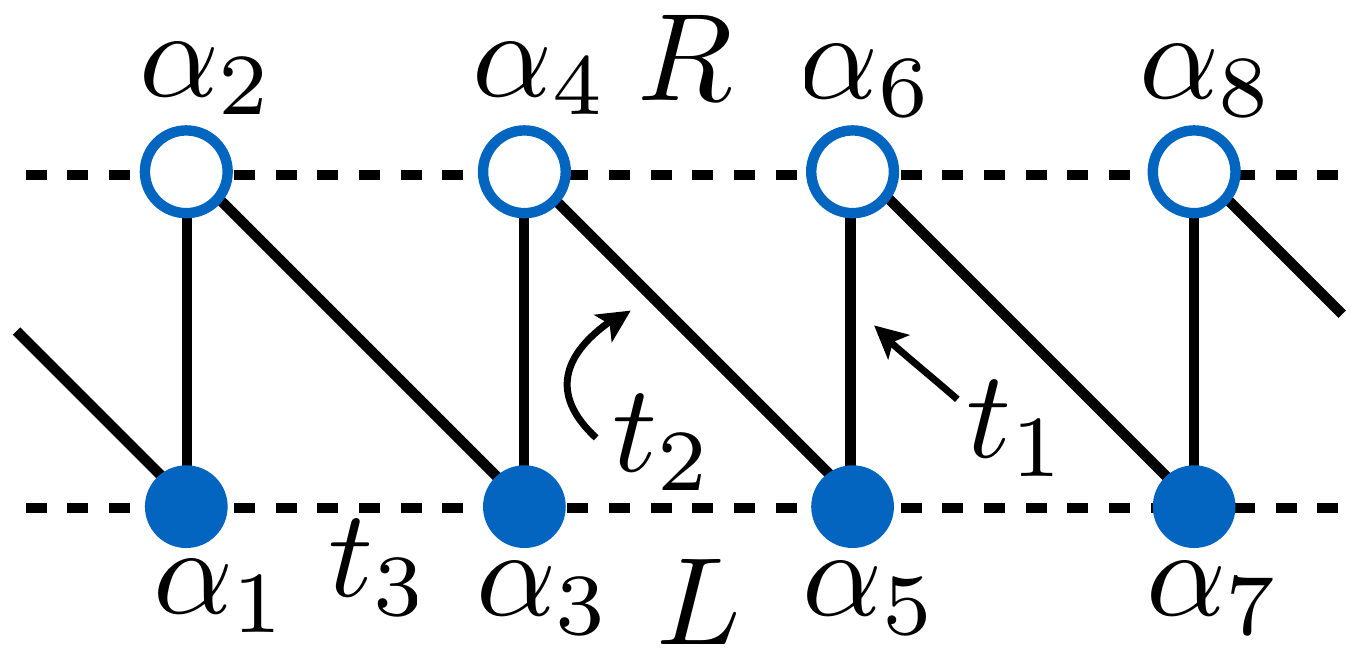}
	\caption{%
		Parafermion ordering path used to discuss the self-duality of the two-leg ladder model studied in Section~\ref{sec:two_chain}. Solid lines and
		numbering of the parafermion operators indicate the ordering path, with coincides with the $t_1$ and $t_2$ interactions. The $t_3$ interactions, shown as dashed lines,
		involve processes that take place above or below the path.}
\label{fig:2leg_path}
\end{figure}

Consider the model of Eq.~\eqref{eqn:model} defined on a two-leg ladder (i.e., $N_y = 2$) with $t_{1,2,3}$ purely real as usual.  In this Appendix we show that the ladder exhibits a variant of self-duality when $t_1 = t_2$, independent of the strength of $t_3$, that sharply constrains the location of the critical line separating the topological and trivial phases found numerically; recall Fig.~\ref{fig:2leg_phase_diagram}.
Following the discussion of Appendix \ref{appendix:dmrg_path}, it will prove useful to equivalently rewrite the Hamiltonian in terms of a \emph{single} chain of parafermions ordered along a path that follows the $t_1$ and $t_2$ bonds.
The specific path, including conventions for $R$ and $L$, appears in Fig.~\ref{fig:2leg_path}.

In this convention the two-leg ladder Hamiltonian becomes
\begin{align}\begin{split}
\label{eq:2LegPFH}
H_{\text{2-leg}}&= -\sum_j\big{[} t_1\omega\alpha_{R,2j}^\dagger\alpha_{R,2j-1}+t_2\omega\alpha_{R,2j+1}^\dagger\alpha_{R,2j} \\
      &+ t_3(\omega\alpha_{R,2j+2}^\dagger \alpha_{R,2j} +\omega^{2}\alpha_{L,2j+1}^\dagger \alpha_{L,2j-1}) + \text{H.c.}\big{]}.
\end{split}\end{align}
The first line coincides with the nearest-neighbor parafermion chain discussed in Sec.~\ref{sec:models} and Appendix~\ref{appendix:Potts_mapping}; the second line simply adds second-neighbor interactions.

Equations~\eqref{eqn:Rmapping} and \eqref{eqn:Lmapping} allow us to alternatively express the Hamiltonian in Potts language:
\begin{align}\begin{split}
  H_{\text{2-leg}} &=-\sum_j\{t_2(\sigma_{j+1}^\dagger\sigma_j + {\rm H.c.}) + t_1(\tau_j+\tau_j^\dagger) \\
  &\;\quad+ t_3 [\omega^2(\tau_j + \tau_{j+1}^\dagger)\sigma_{j+1}^\dagger \sigma_j + {\rm H.c.}]\} .
  \label{H2leg}
\end{split}\end{align}
Next we introduce a duality transformation, defining operators
\begin{equation}
  \mu_j \equiv \prod_{i<j} \tau_j,~~~~~\nu_j \equiv \sigma_{j-1}^\dagger \sigma_j
\end{equation}
that satisfy the same algebra as the original $\sigma,\tau$ variables, e.g., $\mu_j \nu_j = \omega \nu_j \mu_j$.
In this dual representation we obtain
\begin{align}\begin{split}
  H_{\text{2-leg}} &= -\sum_j\{t_1(\mu_{j+1}^\dagger\mu_j + {\rm H.c.}) + t_2(\nu_j+\nu_j^\dagger) \\
  &\quad\; + t_3 [\omega(\nu_j^\dagger + \nu_{j+1})\mu_{j+1}^\dagger \mu_j + {\rm H.c.}]\}.
  \label{Hdual}
\end{split}\end{align}
The first line retains the same form as in Eq.~\eqref{H2leg} but with $t_1$ and $t_2$ swapped.
Crucially, the $t_3$ term in the second line \emph{also} retains the same form upon additionally performing an antiunitary `time-reversal' transformation that sends
\begin{equation}
  \nu_j\rightarrow \nu_j^\dagger,~~~~~\mu_j \rightarrow \mu_j.
\end{equation}
(Note that this transformation preserves the commutation relations for the dual variables.)
Thus, as advertised, $H_{\text{2-leg}}$ is invariant under duality followed by time reversal when $t_1 = t_2$ for arbitrary $t_3$.
This observation constrains the critical line separating the trivial and topological gapped phases for the two-leg ladder to reside precisely along the line $t_1 = t_2$, which is indeed borne out in our numerical simulations (Fig.~\ref{fig:2leg_phase_diagram}).

\section{Scaling argument for Fibonacci phase boundary \label{appendix:CFT}}

With $t_1=t_2=0$ and $t_3>0$ our Hamiltonian in Eq.~\eqref{eqn:model} describes $N_y$ independent parafermion chains, each tuned to the self-dual critical point described by a non-chiral $\Z3$ parafermion CFT.
This CFT contains six primary fields $\{1, \psi, \psi^\dag, \varepsilon, \sigma, \sigma^\dag\}$ with respective scaling dimensions $\{0,\frac23,\frac23,\frac25,\frac1{15},\frac1{15}\}$.
Reference~\onlinecite{Mong:2014b} relates these fields directly to the lattice parafermion operators of our model in the long wavelength limit:
\begin{align}\begin{split}
	\alpha_{R,j}\sim a\psi_R+b(-1)^j\Phi_{\sigma_R\epsilon_L}+\dots, \\
	\alpha_{L,j}\sim a\psi_L+b(-1)^j\Phi_{\epsilon_R\sigma_L}+\dots,
	\label{AF_alpha_expansion2}
\end{split}\end{align}
with non-universal coefficients $a,b$ and $\Phi_{AB}$ denoting the fusion product of fields $A$ and $B$.
The first term in each expansion has a scaling dimension of $\frac{2}{3}$, while the second has a (smaller) scaling dimension of $\frac{7}{15}$.
The ellipses indicate higher-order terms with scaling dimension larger than $1$.

Using Eqs.~\eqref{AF_alpha_expansion2}, the interchain Hamiltonian of Eq.~\eqref{eqn:model} becomes
\begin{align}
	H_\perp &\sim-\sum_j \, \Big[ (t_1+t_2) \omega  a^2\psi_L^\dagger(j,y)\psi_R(j,y+1) \nonumber\\
	&\quad+ (t_1-t_2) \omega b^2\Phi^\dagger_{\epsilon_R\sigma_L}(j,y)\Phi_{\sigma_R\epsilon_L}(j,y+1)+ \text{H.c.}\Big]  \notag\\
	&\quad+\dots
\end{align}
where we have retained only the two most relevant couplings.
The wire construction of the Fibonacci phase in which we are interested\cite{Mong:2014} arises from the $(t_1+t_2)$ term in the above expansion.
When $t_1=t_2>0$, this coupling clearly dominates.
In this case, the left movers in the CFT of one chain couple \emph{only} to the right movers of the next, gapping out the interior modes and leaving a chiral $\Z3$ parafermion CFT at the top and bottom edges.
However, it is important to note that the second term (for which left and right movers of each chain interact, leaving the fate of the system less clear) is actually more relevant:
The first coupling has scaling dimension $4/3$, while the second has dimension $14/15$.

One can use this scaling information to estimate the shape of the phase boundary separating the Fibonacci phase from the state favored by the $(t_1-t_2)$ term.
Defining $\lambda_\pm=t_1\pm t_2$, the renormalization group flow equations for the couplings $\lambda_\pm$ read
\begin{equation}
  \partial_l\lambda_\pm=(2-\Delta_\pm)\lambda_\pm,
\end{equation}
where $l$ is a logarithmic rescaling factor while $\Delta_+=\frac{4}{3}$ and $\Delta_-=\frac{14}{15}$ denote the corresponding scaling dimensions.
By rearranging the flow equation, one can see that the combination $\lambda_+^{(2-\Delta_-)}/\lambda_-^{(2-\Delta_+)}$ does not change with the renormalization flow.
At criticality the $\lambda_+$ and $\lambda_-$ terms compete to a draw.  The phase boundary should therefore reside along a constant value of this flow invariant in order for the two coupling constants to remain in competition as $l$ increases.
In our case, this means that the phase boundary should occur approximately along a curve given by
\begin{equation}
 t_{2c}-t_{1c} = C\,(t_{2c}+t_{1c})^{8/5}
\end{equation}
for some non-universal constant $C$.

\bibliography{dmrg_fibonacci}

\end{document}